\documentclass[amsmath,amssymb,aps,prd,11pt,tightenlines,superscriptaddress,
nofootinbib,preprintnumbers,notitlepage]{revtex4-1}

\usepackage{booktabs,multirow,graphicx,tabularx,mathtools,slashed,url,hyperref}
\usepackage[maxfloats=100]{morefloats}
\usepackage{color,xcolor,braket,comment}
\usepackage[normalem]{ulem}
\usepackage{enumitem}
\usepackage{feynmp}
\usepackage{blkarray,bigstrut}

\makeatletter
\@ifundefined{pdfoutput}{}{\DeclareGraphicsRule{*}{mps}{*}{}}
\def\l@subsubsection#1#2{}
\makeatother

\newcommand{\tnaive}{\hat{T}}

\newcommand\topstrut[1][1.2ex]{\setlength\bigstrutjot{#1}{\bigstrut[t]}}
\newcommand\botstrut[1][0.9ex]{\setlength\bigstrutjot{#1}{\bigstrut[b]}}

\newcommand{\qqquad}{\qquad \qquad}

\setlist[itemize]{itemsep=1pt,parsep=1pt, topsep=3pt}
\setlist[enumerate]{itemsep=1pt,parsep=1pt, topsep=3pt}

\newcommand{\lag}{\mathcal{L}}

\newcommand{\ope}[1]{\mathcal{O}_{#1}}

\newcommand{\eg}{e.\,g.\ }

\newcommand{\ifb}{\ensuremath \mathrm{fb}^{-1}}

\newcommand{\gev}{{\ensuremath \mathrm{GeV}}}

\newcommand{\phisq}{\phi^\dagger \phi}

\newcommand{\pder}[2]{\frac {\partial #1} {\partial #2}}

\DeclareMathOperator{\Imag}{Im}
\DeclareMathOperator{\Pois}{Pois}

\newcommand{\boldx}{\ensuremath \mathbf{x}}

\newcommand{\boldg}{\ensuremath \mathbf{g}}

\newcommand{\arxiv}[1]{\href{http://arxiv.org/abs/#1}{arXiv:#1}}

\AtBeginDocument{
  \heavyrulewidth=.08em
  \lightrulewidth=.05em
  \cmidrulewidth=.03em
  \belowrulesep=.65ex
  \belowbottomsep=0pt
  \aboverulesep=.4ex
  \abovetopsep=0pt
  \cmidrulesep=\doublerulesep
  \cmidrulekern=.5em
  \defaultaddspace= .5em
  \setlength{\tabcolsep}{0.5em}
}

\begin{document}


\title{Better Higgs-CP Tests Through Information Geometry}

\author{Johann Brehmer}
\affiliation{Center for Cosmology \& Particle Physics and Center for Data Science, New York University, USA}

\author{Felix Kling}
\affiliation{Department of Physics and Astronomy, University of California, Irvine, USA}

\author{Tilman Plehn}
\affiliation{Institut f\"ur Theoretische Physik, Universit\"at Heidelberg, Germany}

\author{Tim M.P. Tait}
\affiliation{Department of Physics and Astronomy, University of California, Irvine, USA}
\affiliation{Institute of Physics and Astronomy, University of Amsterdam, The Netherlands}

\preprint{UCI-HEP-TR-2017-14}

\begin{abstract}
  Measuring the $CP$ symmetry in the Higgs sector is one of the key
  tasks of the LHC and a crucial ingredient for precision studies, for
  example in the language of effective Lagrangians. We systematically
  analyze which LHC signatures offer dedicated $CP$ measurements in the
  Higgs-gauge sector, and discuss the nature of the information they
  provide. Based on the Fisher information measure, we compare the
  maximal reach for $CP$-violating effects in weak boson fusion,
  associated $ZH$ production, and Higgs decays into four leptons. We
  find a subtle balance between more theory-independent approaches and
  more powerful analysis channels, indicating that rigorous evidence
  for $CP$ violation in the Higgs-gauge sector will likely require a
  multi-step process.
\end{abstract}

\maketitle
\tableofcontents
\begin{fmffile}{feynman}
\newpage

\section{Introduction}
\label{sec:intro}

Since the experimental observation of the Higgs boson at the Large
Hadron Collider (LHC)~\cite{higgs,discovery}, detailed studies of its
properties have become one of the most important laboratories to
search for physics beyond the Standard Model.  With the measurement of
the Higgs mass, the last remaining parameter of the Standard Model has
been determined. This implies that further Higgs measurements can be
viewed as consistency checks on the validity of the Standard Model
description.  In particular, deviations from the Standard Model
expectations induced by heavy new particles can be described by a
continuous and high-dimensional parameter space of Wilson coefficients
in the Lagrangian of an effective field theory
(EFT)~\cite{eftfoundations, eftorig, eftreviews, mike_ilaria}.  EFT
descriptions have the advantage that they are well-defined quantum
field theories and allow us to predict and include kinematic
distributions in the analysis~\cite{higgs_fit,yr4}.

The key assumptions defining any effective Lagrangian are the particle
content and the symmetry structure. Once these two initial assumptions
are agreed upon, the Lagrangian is defined as a power series in the
heavy new physics scale $\Lambda$.  First, the general consensus is
that the particle content of Higgs analyses is given by the Standard
Model particles~\cite{our750}.  However, on the symmetry side, the
situation is less clear. To begin with, one can embed the Higgs scalar
in a SM-like $SU(2)_L$ doublet or add a scalar field unrelated to the
Goldstone modes. In this paper we realize the electroweak gauge
symmetry linearly and include a complex Higgs-Goldstone doublet.  A
remaining question concerns the charge conjugation ($C$) and parity
($P$) symmetries of the Higgs boson and its interactions.  In the
Standard Model, after the CKM rotations which diagonalize the fermion
masses, the Higgs boson has $C$ and $P$ preserving interactions at
tree level.  Any deviation from this prediction would be a striking
manifestation of physics beyond the Standard Model, and it is
experimentally exigent to determine whether there are new sources of
$CP$ violation in the Higgs sector.

A common approach addresses this question by simplistically combining
$CP$-even and $CP$-odd operators into one effective Lagrangian and
fitting them to a combination of arbitrary observables. Because of
many caveats affecting global dimension-six EFT analyses, the results of
such an analysis do not say much about the $CP$ nature of the Higgs
boson.  Instead, we propose to carefully disentangle three
questions~\cite{tao1,tao2}:
\begin{enumerate}
\item Which LHC observables are sensitive to the $CP$ nature of the
  Higgs boson?
\item What are the assumptions linking these observables to $CP$?
\item How well can we quantitatively test the Higgs' $CP$ properties
  based on these observables?
\end{enumerate}
Once such a dedicated analysis establishes that $CP$ is not a good
symmetry of the Higgs sector, we will expand the effective
Lagrangian to include $CP$-violating operators to better discern the
nature of the $CP$ violation.  The first two questions have
straightforward answers~\cite{german_review}. In fact, there exists a
wealth of individual LHC studies for this kind of measurement in the
Higgs-gauge~\cite{cabibbo, nelson, phi_jj, phi_jjs, cp_zh, cp_dec, hopkins,
  cp_fun, spanno, our_nelson} and in the Yukawa sectors~\cite{cp_gg, cp_tth,
  cp_tau}, as well as through global analyses~\cite{cp_global}. Our
focus therefore lies on a unified theory framework which allows us to
systematically compare and combine the leading three Higgs-gauge LHC
production/decay channels.

Progress on the third question requires the use of modern analysis and
tools.  The LHC experiments have come to rely on high-level
statistical discriminants, including hypothesis tests based on
multivariate analysis with machine learning or the matrix element
method~\cite{matrix_element, kyle_review}.  These tools are able to
tease out features that defy simpler cut-and-count analysis based on
one-dimensional or two-dimensional kinematic distributions.  We apply
the new \textsc{MadFisher} approach~\cite{madfisher} based on
information geometry~\cite{information-geometry} to systematically
study the sensitivity of different Higgs processes to different
scenarios of $CP$ violation.  Through the Cram\'er-Rao bound, the
Fisher information determines the maximum knowledge about model
parameters that can be derived from a given
experiment~\cite{cramer-rao}.  It allows us to define and to compute
the best possible outcome of any multivariate black-box
analysis~\cite{kyle_review, madmax1} as well as the expected outcome
based on a more limited set of kinematic observables. In this way, we
determine not only which Higgs production and decay processes are
best-suited to test its $CP$ properties, but also identify which
kinematic variables carry the relevant information.\bigskip

We begin with a brief review of $CP$-sensitive observables at the LHC
in Sec.~\ref{sec:intro_cp_lhc}, $CP$ violation in the Higgs-gauge
sector in Sec.~\ref{sec:intro_cp_higgs}, and our Fisher information
approach in Sec.~\ref{sec:intro_formalism}. We study the three leading
LHC signatures, Higgs production in weak boson fusion (WBF) in
Sec.~\ref{sec:wbf}, associated $ZH$ production in Sec.~\ref{sec:zh},
and Higgs decays to four leptons in Sec.~\ref{sec:dec}. For each of
these signatures we discuss the possible $CP$-sensitive observables
and briefly describe the advantages and challenges of the
corresponding LHC analysis. In Sec.~\ref{sec:summary}, we compare 
all three channels.

\subsection{CP vs naive time reversal}
\label{sec:intro_cp_naive}

As is well known, the three discrete symmetries consistent with
Lorentz invariance and a Hamiltonian which is
Hermitian~\cite{german_review} are charge conjugation ($C$), parity
($P$), and time reversal ($T$).  These three operators act on a
complex scalar field $\phi(t,\vec x)$ as:
\begin{align}
 C \, \phi(t, \vec x) \, C^{-1} = \eta_C \; \phi^*(t, \vec x) \qquad 
 P \, \phi(t,\vec x) \, P^{-1}  = \eta_P \; \phi(t, -\vec x) \qquad
 T \, \phi(t, \vec x) \, T^{-1} = \eta_T \; \phi(-t, \vec x),
\end{align}
where the phases $\eta_j$ define the intrinsic symmetry properties of $\phi$. 
$C$ and $P$ are unitary transformations, while $T$ is anti-unitary, implying that the phase
$\eta_T$ is not measurable and can be chosen to be $\eta_T = 1$.
Acting on a single-particle state with 4-momentum $p$ and spin $s$ produces
\begin{align}
  C \ket{\phi (p,s)} = \ket{\phi^* (p, s)} \qquad 
  P \ket{\phi (p,s)} = \eta_\phi \ket{\phi(-p, s)} \qquad 
  T \ket{\phi (p,s)} = \bra{\phi (-p, -s)} \; ,
\label{eq:dev_cpt}
\end{align}
where $\eta_\phi$ is the intrinsic parity of the field.  Time reversal
transforms incoming states into outgoing states, so it is convenient
to define a `naive time reversal'~\cite{german_review,tao1,choudhury}
\begin{align}
  \tnaive \ket{\phi (p, s)} = \ket{\phi (-p, -s)} \; ,
\label{eq:def_that}
\end{align}
which explicitly omits exchanging initial and final states.\bigskip

Observables can be chosen to reflect the $C$, $P$, or $T$
transformation properties of the underlying transition amplitude.  We
are interested in a real-valued observable $O$ that can be measured in
a process $\ket{i} \to \ket{f}$. Interesting observables at the LHC
are functions of the 4-momenta, spins, flavors, and charges of initial
and final states. First, we define a $U$-odd or $U$-even observable as
\begin{align}
O ( U \ket{i} \to U \ket{f} ) = \mp \,  O \, ( \ket{i} \to \ket{f} ) 
\qqquad \text{for} \quad U=C,P,\tnaive \; . 
\label{eq:def_odd}
\end{align}
where the upper (lower) sign refers to $U$-odd ($U$-even).\bigskip

For the purpose of testing the properties of the underlying theory, a
\emph{genuine} $U$-odd observable is defined as having a vanishing
expectation value in a $U$-symmetric theory (for which
$\lag = U \lag U^{-1}$),
\begin{align}
  \braket{O}_{\lag = U \lag U^{-1}} = 0 \; .
\label{eq:genuine1}
\end{align}
In case the initial state is a $U$ eigenstate, or the probability
distribution of the initial states $p(\ket{i})$ is $U$-symmetric, the
second definition is slightly weaker. One can show that under this
condition any $U$-odd observable is also genuinely $U$-odd,
\begin{align}
O ( U \ket{i} \to U \ket{f} ) = - \,  O \, ( \ket{i} \to \ket{f} ) \quad \text{(odd)} 
\;\;
\stackrel{
p( \ket{i} ) = p ( U\ket{i} )
}{\Longrightarrow} 
\; \;
\braket{O}_{\lag =U \lag U^{-1}} = 0 \quad \text{(genuine odd)} \; .
\label{eq:odd_relation}
\end{align}
In particular, any observable that compares the probabilities of two
conjugated processes
\begin{align}
  O \propto \mathrm{d} \sigma ( \ket{i} \to \ket{f} ) - \mathrm{d}  \sigma ( U\ket{i} \to U\ket{f} )
\label{eq:genuine2}
\end{align}
is obviously genuinely $U$-odd.
\bigskip

We can gain additional insights on $CP$ from the $\tnaive$
transformation properties.  Based on the definition in
Eq.\;\eqref{eq:genuine1}, at tree level, a finite expectation value of
a genuine $\tnaive$-odd observable $O$ indicates a $CP$-violating
theory~\cite{atwood}. In addition to $CPT$ invariance this argument
requires:
\begin{itemize}
\item the phase space is $\tnaive$-symmetric;
\item the initial state is a $\tnaive$-eigenstate, or its distribution is invariant under $\tnaive$; and
\item there cannot be re-scattering effects.
\end{itemize}
The latter correspond to absorptive, complex-valued loop contributions, for instance an imaginary part in the propagator of 
an intermediate on-shell particle. To illustrate this point,
consider the transition amplitude $\mathcal{T}$ defined via $S = 1 + i \mathcal{T}$. 
The matrix elements satisfy
\begin{align}
\langle f | \mathcal{T} | i \rangle 
 \stackrel{\text{$CP$-invariant}}{=} \langle i_T | \mathcal{T} | f_T \rangle 
 \stackrel{\text{no re-scattering}}{=} \langle f_T | \mathcal{T} | i_T \rangle^* 
\qquad \Rightarrow \qquad 
|\langle f | \mathcal{T} | i
\rangle |^2 = |\langle f_T | \mathcal{T} | i_T \rangle |^2 \; ,
\end{align}
with $T\ket{i}=\langle i_T|$ as given in Eq.\;\eqref{eq:dev_cpt}, etc.
The first step follows from $CPT$ invariance, and the second from the
optical theorem in the absence of re-scattering. Indeed, in a
$CP$-symmetric theory and in the absence of re-scattering, the matrix
element squared is $\tnaive$-invariant.\bigskip

In practice, this argument means that where genuine $CP$ observables
cannot be constructed, we can analyze genuine $\tnaive$ observables
instead. A non-zero expectation value here is evidence for $CP$
violation under the additional assumption of no or negligible
re-scattering.

\subsection{CP violation in LHC processes}
\label{sec:intro_cp_lhc}

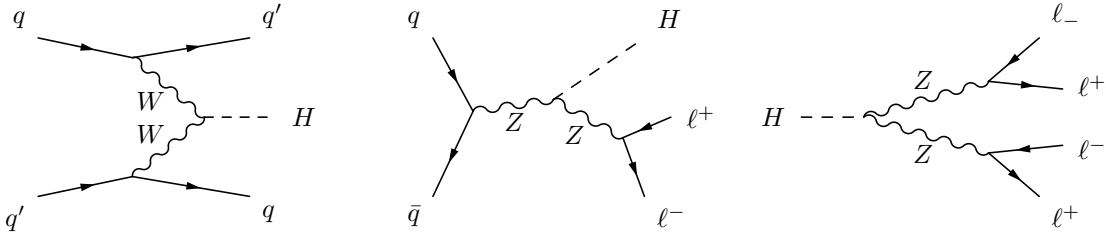
\begin{figure}[t]
  \begin{center}
    \begin{fmfgraph*}(100,60)
      \fmfset{arrow_len}{2mm}
      \fmfleft{i2,i1}
      \fmfright{o3,o2,o1}
      \fmf{fermion,width=0.6,lab.side=left,tension=2}{i1,v1}
      \fmf{fermion,width=0.6,lab.side=left,tension=1}{v1,o1}
      \fmf{photon,width=0.6,lab.side=right,label=$W$,label.dist=2,tension=1}{v1,v3}
      \fmf{dashes,width=0.6,lab.side=left,tension=2}{v3,o2}
      \fmf{photon,width=0.6,lab.side=right,label=$W$,label.dist=2,tension=1}{v3,v2}
      \fmf{fermion,width=0.6,lab.side=left,tension=2}{i2,v2}
      \fmf{fermion,width=0.6,lab.side=left,tension=1}{v2,o3}
      \fmflabel{$q$}{i1}
      \fmflabel{$q'$}{i2}
      \fmflabel{$q'$}{o1}
      \fmflabel{$q$}{o3}
      \fmflabel{$H$}{o2}
    \end{fmfgraph*}
    \hspace*{15mm}
    \begin{fmfgraph*}(100,60)
      \fmfset{arrow_len}{2mm}
      \fmfleft{i2,i1}
      \fmfright{o3,o2,o1}
      \fmf{fermion,width=0.6,lab.side=left,tension=2}{i1,v1}
      \fmf{fermion,width=0.6,lab.side=left,tension=2}{v1,i2}
      \fmf{photon,width=0.6,lab.side=right,label=$Z$,label.dist=3,tension=2}{v1,v2}
      \fmf{dashes,width=0.6,lab.side=left,tension=1}{v2,o1}
      \fmf{photon,width=0.6,lab.side=right,label=$Z$,label.dist=3,tension=1}{v2,v3}
      \fmf{fermion,width=0.6,lab.side=left,tension=1}{o2,v3}
      \fmf{fermion,width=0.6,lab.side=left,tension=1}{v3,o3}
      \fmflabel{$q$}{i1}
      \fmflabel{$\bar{q}$}{i2}
      \fmflabel{$\ell^+$}{o2}
      \fmflabel{$\ell^-$}{o3}
      \fmflabel{$H$}{o1}
    \end{fmfgraph*}
    \hspace*{15mm}
    \begin{fmfgraph*}(100,60)
      \fmfset{arrow_len}{2mm}
      \fmfleft{i1}
      \fmfright{o4,o3,o2,o1}
      \fmf{dashes,width=0.6,lab.side=left,tension=4}{i1,v1}
      \fmf{photon,width=0.6,lab.side=left,label=$Z$,label.dist=4,tension=1}{v1,v2}
      \fmf{fermion,width=0.6,lab.side=left,tension=1}{o1,v2}
      \fmf{fermion,width=0.6,lab.side=left,tension=1}{v2,o2}
      \fmf{photon,width=0.6,lab.side=left,label=$Z$,label.dist=4,tension=1}{v3,v1}
      \fmf{fermion,width=0.6,lab.side=left,tension=1}{o3,v3}
      \fmf{fermion,width=0.6,lab.side=left,tension=1}{v3,o4}
      \fmflabel{$H$}{i1}
      \fmflabel{$\ell_-$}{o1}
      \fmflabel{$\ell^+$}{o2}
      \fmflabel{$\ell^-$}{o3}
      \fmflabel{$\ell^+$}{o4}
    \end{fmfgraph*}
  \end{center}
  \caption{Feynman diagrams describing the three processes considered in
    this paper: WBF Higgs production, associated $ZH$ production, and $H
    \to 4 \ell$ decays.}
  \label{fig:feyn}
\end{figure}

We evaluate the effect of $CP$-odd operators on the three most
promising LHC Higgs signatures: WBF Higgs production in
Sec.~\ref{sec:wbf}, associated $ZH$ production in Sec.~\ref{sec:zh},
and Higgs decays to four leptons in Sec.~\ref{sec:dec}.  From
Fig.~\ref{fig:feyn}, it is clear that these three processes are
governed by the same hard process, with different initial and final
state assignments, and the $W$ and $Z$ couplings related by custodial
symmetry. 

\begin{table}[t!]
\centering
  \begin{tabular}{lccccr}
  \toprule 
  Observable & Theory & Re-scattering  & Symmetry argument && Prediction \\
  \midrule
  \multirow{4}{*}{$CP$-odd, $\tnaive$-odd} & \multirow{2}{*}{$CP$-symmetric} & no & $CP$ and $\tnaive$: symmetric $\sigma_\text{int}$, odd $O$ & $\Rightarrow$& $\langle O \rangle=0$ \\ 
  & & yes &  $CP$: symmetric $\sigma_\text{int}$, odd $O$ & $\Rightarrow$& $\langle O \rangle=0$\\ 
  \cmidrule{2-6}
  & \multirow{2}{*}{$CP$-violating} & no & \multicolumn{3}{r}{can have $\langle O \rangle \neq 0$} \\ 
  & & yes & \multicolumn{3}{r}{can have $\langle O \rangle \neq 0$}  \\ 
  \midrule
  \multirow{4}{*}{$CP$-odd, $\tnaive$-even} & \multirow{2}{*}{$CP$-symmetric} & no &  $CP$: symmetric $\sigma_\text{int}$, odd $O$ & $\Rightarrow$& $\langle O \rangle=0$\\ 
  & & yes &  $CP$: symmetric $\sigma_\text{int}$, odd $O$ & $\Rightarrow$& $\langle O \rangle=0$\\ 
  \cmidrule{2-6}
  & \multirow{2}{*}{$CP$-violating} & no &  $\tnaive$: anti-symmetric $\sigma_\text{int}$, even $O$& $\Rightarrow$& $\langle O \rangle=0$\\ 
  & & yes & \multicolumn{3}{r}{can have $\langle O \rangle \neq 0$} \\ 
  \bottomrule
  \end{tabular}
  \caption{Predictions for $CP$-odd observables $O$ based on the
    theory's symmetries and the observable's transformation properties
    under $\tnaive$. In all cases we assume that the initial state or
    its probability distribution is symmetric under both $CP$ and
    $\tnaive$. 
    }
\label{tab:tn-odd}
\end{table}

 Since it is not realistically possible to determine the spins in the
 initial or final states in these processes, all observables must be
 constructed as functions of the 4-momenta. Ideally, we can
 reconstruct four independent external 4-momenta for each process
 shown in Fig.~\ref{fig:feyn} and combine them into ten scalar products of the
 type $(p_i \cdot p_j)$. Four of them correspond to the masses of the
 initial- and final-state particles, and the remaining six specify the
 kinematics.  Scalar products are $P$-even.
 Equation\;\eqref{eq:def_that} implies that they transform the same
 way under $\tnaive$ as under $P$, so they are also $\tnaive$-even.
 As we will see in Sec.~\ref{sec:wbf_obs}, two of them are $C$-odd,
 while the remaining four are $C$-even.  In addition to the scalar
 products, there is one $P$-odd and $\tnaive$-odd observable
 constructed from four independent 4-momenta,
 $\epsilon_{\mu\nu\rho\sigma} k_1^\mu k_2^\nu q_1^\rho
 q_2^\sigma$~\cite{phi_jj, phi_jjs}.  Altogether, there are:
\begin{itemize}[label=\raisebox{0.1ex}{\scriptsize$\bullet$}]
\item four scalar products corresponding to  masses of the external particles;
\item four $C$-even, $P$-even, and $\tnaive$-even scalar products;
\item two $C$-odd, $P$-even, and $\tnaive$-even scalar products;
\item one $C$-even, $P$-odd, and $\tnaive$-odd observable constructed from the Levi-Civita-tensor,
\end{itemize}
for all three processes illustrated in Fig.~\ref{fig:feyn}. More
details, including some analytic results, are given in the
appendix. Thus, at most three observables are $CP$-odd, with 
the main difference between each process coming from what can be measured 
for each of the four fermion lines.\bigskip

In cases where the initial state is guaranteed to be $CP$-even and
$\tnaive$-even, or can be boosted into such a frame, we can
distinguish two types of $CP$-odd observables:
\begin{enumerate}
\item $CP$-odd and $\tnaive$-odd: for the $q\bar{q}$ initial state
  this implies that the observable is also genuine $CP$-odd and genuine
  $\tnaive$-odd (see Eq.\;\eqref{eq:odd_relation}).  In a $CP$-symmetric
  theory its expectation value vanishes, implying that a non-zero
  expectation value requires $CP$ violation regardless of the presence of re-scattering. 
  The different cases are illustrate in the
  upper half of Table~\ref{tab:tn-odd}.
\item $CP$-odd and $\tnaive$-even: for the $q\bar{q}$ initial state
  the observable is also genuine $CP$-odd, so in a $CP$-symmetric
  theory its expectation value vanishes. In the lower half of
  Table~\ref{tab:tn-odd} we show the different scenarios: if the
  theory is $CP$-violating, the corresponding expectation values does
  not vanish.  If we ignore re-scattering, the theory also appears
  $\tnaive$-violating, but the expectation value of the $\tnaive$-even
  observable combined with an anti-symmetric amplitude will still
  vanish.  However, in the presence of re-scattering or another
  complex phase, this unwanted condition from the $\tnaive$ symmetry
  vanishes, and the expectation for $\braket{O}$ matches the symmetry
  of the theory.
\end{enumerate}
This implies that for a (statistically) $CP$-symmetric initial state, one can arrive at a
meaningful statement about the $CP$ symmetry of the underlying theory
either through a $CP$-odd and $\tnaive$-odd observable without any
assumption about complex phases or through a $CP$-odd and
$\tnaive$-even observable in the presence of an complex phase.

\subsection{CP violation in the Higgs-gauge sector}
\label{sec:intro_cp_higgs}

Typical tests of $C$, $P$, or $T$ symmetries of the Higgs sector
do not probe the symmetry nature of the actual Higgs
field, but rather the transformation properties of the
action through its influence on $S$-matrix elements.  We focus on the
transformation properties of observables and explore how they reflect the
symmetry structure of the Higgs Lagrangian.
To this end, we evaluate the effect of $CP$-violating as opposed to
$CP$-conserving Higgs couplings to weak bosons or heavy fermions. For
an effective Higgs-gauge Lagrangian truncated at mass dimension six,
\begin{align}
  \lag = \lag_\text{SM} + \frac {f_i} {\Lambda^2} \ope{i} 
\label{eq:def_wilson}
\end{align}
our $CP$-even reference scenario consists of the renormalizable Standard Model
Lagrangian combined with the five $CP$-even dimension-six operators in
the HISZ basis~\cite{hisz,mike_ilaria,higgs_fit},
\begin{alignat}{2}
  \ope{B}  &= i \frac{g}{2} \, (D^\mu\phi^\dagger) (D^\nu\phi) \, B_{\mu\nu} \quad & \quad
  \ope{W}  &= i \frac{g}{2} \, (D^\mu\phi)^\dagger \sigma^k ( D^\nu\phi) \, W_{\mu\nu}^k \notag \\
  \ope{BB}  &= -\frac{g'^2}{4}  \,  (\phisq) \, B_{\mu\nu} \, B^{\mu\nu} \quad & \quad
  \ope{WW}  &= -\frac{g^2}{4} \, (\phisq) \, W^k_{\mu\nu} \, W^{\mu\nu\, k} \notag \\
  \ope{\phi,2}  &= \frac{1}{2} \, \partial^\mu(\phi^\dagger\phi) \, \partial_\mu(\phi^\dagger\phi) \; .
\label{eq:op_cpeven}
\end{alignat}
At the same mass dimension, $CP$-odd couplings are described by
operators
\begin{align}
\ope{B\tilde{B}}  
&= -\frac{g'^2}{4} \, (\phisq) \, \widetilde{B}_{\mu\nu} \, B^{\mu\nu} 
\equiv -\frac{g'^2}{4} \, (\phisq) \, \epsilon_{\mu \nu \rho \sigma} B^{\rho\sigma} \, B^{\mu\nu} \notag \\
\ope{W\widetilde{W}}  
&= -\frac{g^2}{4} \, (\phisq) \, \widetilde{W}^k_{\mu\nu} \, W^{\mu\nu\, k} 
\equiv -\frac{g^2}{4} \, (\phisq) \, \epsilon_{\mu \nu \rho \sigma} W^{\rho\sigma\, k} \, W^{\mu\nu\, k} \; .
\label{eq:op_cpodd}
\end{align}
With the Levi-Civita tensor, these operators break
down as $C$-conserving and $P$-violating.\bigskip

While the effective Lagrangians in Eqs.\;\eqref{eq:op_cpeven} and
\eqref{eq:op_cpodd} demand real coefficients $f_{WW}$ and
$f_{W\widetilde{W}}$, it is also interesting to observe what happens
when they are taken to be complex.  Strictly speaking, this does not
occur in an EFT from integrating out massive degrees of freedom in a
well-defined UV theory.  However, absorptive complex phases can appear
through light degrees of freedom. Such cases are not technically
described by a local EFT and could lead to different momentum
dependences, so we leave a more refined treatment of this case for
future work. Instead, we consider coefficients such as $f_{WW}$ and
$f_{W\widetilde{W}}$ to be complex to illustrate how such cases
complicate the determination of the $CP$ nature of the Higgs
interactions.  Such complex phases already occur in the Standard
Model, for instance from electroweak corrections or in Higgs
production with a hard jet~\cite{absorptive_higgs}. Such loop-induced
contributions to the expectation value of $CP$-odd observables must be
taken into account in precision measurements.\bigskip

Combining the different pieces, we arrive at thirteen model parameters
of interest,
\begin{align}
  \boldg =  \frac {v^2} {\Lambda^2} 
\big(
     f_{\phi,2} ~ f_W ~ f_B ~ f_{WW} ~ f_{BB} ~ f_{W\widetilde{W}} ~ f_{B\tilde{B}} ~
     \Imag f_{W} ~ \Imag f_{B} ~
     \Imag f_{WW} ~ \Imag f_{BB} ~ \Imag f_{W\widetilde{W}} ~ \Imag f_{B\tilde{B}}
\big)^T \,, \label{eq:gspace}
\end{align}
where the factor $v^2$ ensures that the model parameters are
dimensionless.  The first seven entries represent the usual Wilson
coefficients in the EFT.  The last six entries allow for absorptive
contributions. We will use this full vector of model parameters 
to analyze the sensitivity of different processes to the $CP$
properties of the Higgs-gauge sector.

\subsection{Information geometry and Cram\'er-Rao bound}
\label{sec:intro_formalism}

We briefly review the basics of information geometry applied to Higgs
physics at the LHC, as introduced in Ref.~\cite{madfisher}. The LHC
measurements are represented by a set of events with kinematic
observables $\boldx$.  Their distribution depends on a vector of model
parameters, for example Higgs couplings, with unknown true values
$\boldg$. An analysis leads to an estimator $\hat{\boldg}$, designed
to follow a probability distribution around the true values. For an
unbiased estimator the corresponding expectation values are equal to
the true values,
$\bar{g}_i \equiv E \left[ \hat{g}_i \middle | \boldg \right] = g_i$.
The typical error of the measurement is described by the covariance
matrix
\begin{align}
  C_{ij}(\boldg)
  \equiv E \left[ (\hat{g}_i-\bar{g}_i) (\hat{g}_j-\bar{g}_j)  \middle | \boldg \right] \, ,
  \label{eq:def_cov}
\end{align}
which, as a generalization of the variance in one dimension, gives the
precision of the measurement: the smaller $C_{ij}$, the better one
can measure the combination of couplings $g_i$ and $g_j$.

The second object of interest is the Fisher information matrix, the
first term in a Taylor series of the log-likelihood around its
maximum, which measures the sensitivity of the likelihood of
experimental outcomes $\boldx$ to the model parameters $\boldg$. The
Fisher information matrix can be computed from the probability
distribution $f(\boldx |\boldg)$ for a specific phase space
configuration given a model, as
\begin{align}
  I_{ij}(\boldg)
     \equiv 
      - E \left[
      \frac {\partial^2 \log f(\boldx |\boldg) } {\partial g_i \, \partial g_j}  \middle | \boldg   \right] \; .
  \label{eq:fisher_information}
\end{align}
A large entry in the Fisher matrix implies that the measurement is
particularly sensitive to a given model parameter combination
$g_{i,j}$. Conversely, an eigenvector of the Fisher matrix with zero
eigenvalue indicates a blind direction, corresponding to a combination
of measurements with no expected impact.\bigskip

The Cram\'er-Rao bound~\cite{cramer-rao} links these two tracers of
the sensitivity of a measurement: the (inverse) Fisher information
tells us how much information a given experiment can optimally extract
about a set of model parameters. The covariance matrix gives the
actual uncertainty of the measurements, and its minimum value must be
larger than the inverse Fisher information,
\begin{align}
  C_{ij}(\boldg) \geq (I^{-1})_{ij}(\boldg) \; .
\label{eq:cramer_rao}
\end{align}
The Fisher information is invariant under a reparametrization
of the observables $\boldx$, and transforms covariantly under a
reparametrization of the model parameters $\boldg$. After removing
blind directions, the Fisher information is a symmetric and positive
definite rank-two tensor and defines a metric on the model
space~\cite{information-geometry}. The model-space distance measure
\begin{align}
d ( \boldg_b; \boldg_a ) 
&= \sqrt{(\boldg_a - \boldg_b)_i \, I_{ij}(\boldg_a) \, (\boldg_a - \boldg_b)_j} \, ,
\label{eq:distances}
\end{align}
gives contours of constant distances as optimal error ellipsoids.
Strictly speaking, it is defined in the tangent space at $\boldg_a$,
but can easily be extended to distances calculated along geodesics on
the theory manifold~\cite{madfisher}.  Such local or global distances
track how (un-)likely it is to measure $\hat{\boldg} = \boldg_b$ given
$\boldg = \boldg_a$. In the Gaussian limit the distance value is
measured in standard deviations.\bigskip

The distributions $f(\boldx|\boldg)$ entering
Eq.\;\eqref{eq:fisher_information} can be computed for any model from
Monte-Carlo simulations combined with a detector simulation.  The
corresponding measurement consists of an observed $n$ events
distributed over phase space positions $\boldx$.  For a total cross
section $\sigma(\boldg)$ and an integrated luminosity $L$ the full
probability distribution in Eq.\;\eqref{eq:fisher_information}
factorizes~\cite{kyle_review,madmax1} as
\begin{align}
  f(\boldx_1, \dots, \boldx_n | \boldg) = \Pois ( n | L \sigma(\boldg) ) \; \prod_{i=1}^n f^{(1)} (\boldx_i|\boldg) \, ,
\end{align}
where $f^{(1)} (\boldx|\boldg)$ is the normalized probability
distribution for a single event populating $\boldx$ and can be
computed by standard event generators. The factorized Fisher
information is
\begin{align}
  I_{ij} 
  &= \frac{L}{\sigma} \; \pder {\sigma}{g_i}  \, \pder {\sigma}{g_j}
  - L \, \sigma \; 
    E \left[ \frac{\partial^2 \log f^{(1)}(\boldx|\boldg)}{\partial g_i \, \partial g_j} \right] \; .
\label{eq:fisher_rates}
\end{align}
This total Fisher information can be calculated from Monte-Carlo
simulations.  It defines the best possible precision with which the
parameters $\boldg$ can be measured based on the full observable
space, independent of the (multivariate) analysis strategy. It also
intrinsically includes all directions in theory space and all
correlations between different parameters and does not require any
discretization of the parameter space.\bigskip

Given the discussion in Sec.~\ref{sec:intro_cp_lhc}, an interesting
question is how much of the full information is included in particular
kinematic distributions. To answer it, we alternatively calculate the
information in one-dimensional or two-dimensional histograms of
kinematic observables. This gives the maximum precision with which
parameters can be measured by analyzing a given set of
observables. Comparing this reduced Fisher information to the total
information based on the full phase space lets us quantitatively
analyze whether the clearer theory interpretation of well-defined
$CP$ observables is worth the loss in sensitivity compared to a
multivariate approach.

We evaluate the resulting Fisher information matrices in three ways.
First, we calculate curves of constant distances given by
Eq.\;\eqref{eq:distances} in the space of dimension-six Wilson
coefficients, corresponding to optimal expected exclusion limits of an
analysis. This allows us to study correlations between different
Wilson coefficients, for example between $CP$-violating and
$CP$-conserving operators.

Second, we can rotate the symmetric Fisher information matrix $I_{ij}$
into its diagonal form, defining eigenvectors as a superposition of
model parameters and the corresponding information eigenvalue. In the
diagonal form, the Cram\'er-Rao bound conveniently defines the reach
of a given analysis in each eigenvector direction. Numerically, this
reach can be expressed in terms of the Wilson coefficients
$\Lambda/\sqrt{f}$, as defined in
Eq.\;\eqref{eq:def_wilson}~\cite{madfisher}. We discuss this analysis
in terms of model-space eigenvectors and their reach in
Sec.~\ref{sec:summary}.

Finally, if we are especially interested in a subset of parameters, we
can compute the corresponding Fisher information either setting all
operators to zero, or by profiling over all other operators, as
discussed in detail in the appendix of Ref.~\cite{madfisher}.  In
Sec.~\ref{sec:summary} we use this procedure to analyze the robustness
of signatures from $CP$-violating operators to other scenarios of new
physics.

\section{Higgs production in weak boson fusion}
\label{sec:wbf}

To construct appropriate kinematic observables for WBF Higgs
production, we can in principle make use of three final-state momenta
and two initial-state momenta, where one momentum is linearly
dependent on the other four due to energy-momentum conservation.  We
assume the Higgs decay $H \to \tau \tau$, which allows us to
approximately reconstruct the Higgs 4-momentum~\cite{wbf_tau}, though
this specific choice of decay mode is expected to have little if any
impact on the final results~\cite{phi_jj}.  Throughout the discussion,
we rely on the reconstruction of the Higgs momentum to reconstruct the
missing information about the initial parton momenta.

\subsection{CP observables}
\label{sec:wbf_obs}

The partonic $qq'$ initial states of weak boson fusion is not a $C$
eigenstate.  The discussion in Sec.~\ref{sec:intro_cp_lhc} thus
implies that one cannot construct a production-side genuine
$CP$-sensitive observable in WBF Higgs production. On the other hand,
Eq.\;\eqref{eq:def_that} states that in the absence of spin
information the transformation properties under $P$ and $\tnaive$ are
the same, and thus one can construct exactly one genuine $\tnaive$-odd
observable based on the Levi-Civita tensor in the center-of-mass
frame, making use of the fact that the initial state probability
distribution is $\tnaive$-symmetric in proton-proton collisions.  In
the absence of large re-scattering effects, it probes $CP$ violation
in the Higgs-gauge sector.  This observable can be naively defined
as~\cite{phi_jj,phi_jjs}
\begin{align}
\epsilon_{\mu\nu\rho\sigma} \; k_1^\mu~ k_2^\nu ~q_1^\rho ~q_2^\sigma \; ,
\label{eq:p_odd_wbf1}
\end{align}
where the two incoming parton momenta are $k_{1,2}$ and the two
outgoing tagging jet momenta are $q_{1,2}$.
However, this definition suffers from the feature that it
changes sign under exchange of the two tagging jet momenta $q_1 \leftrightarrow q_2$.
We remove this ambiguity through the modification~\cite{tao1}
\begin{align}
O \equiv \epsilon_{\mu\nu\rho\sigma} \; k_1^\mu ~k_2^\nu ~q_1^\rho ~q_2^\sigma
    \; \text{sign} \left[ (k_1-k_2)\cdot (q_1-q_2) \right] \; .
\label{eq:p_odd_wbf2}
\end{align}
Defining $k_+$ and $k_-$ to be the initial state momenta in the lab
frame pointing along the positive and negative beam axis
($z$-direction), $q_+$ and $q_-$ are delineated `forward' and
`backward' such that $k_+$ and $q_+$ point to the same hemisphere, or
more generally $(q_+-q_-)\cdot(k_+-k_-) > 0$~\cite{phi_jjs}.  This
implies that $(q_+)_z > (q_-)_z$ in the center-of-mass frame.  In this
notation,
\begin{align}
O = \epsilon_{\mu\nu\rho\sigma} \; k_+^\mu k_-^\nu q_+^\rho q_-^\sigma \; .
\label{eq:p_odd_wbf3}
\end{align}
\bigskip

In the laboratory frame, $k_\pm = (E_\pm,0,0, \pm E_\pm)$.
The assignment for $q_\pm$ imply that the sign factor is
always unity, which reduces $O$ to a triple product
\begin{align}
O 
&=  2 E_+ E_- (  q_{y,+}  q_{x,-}-q_{x,+}  q_{y,-}   )
=  2 E_-  (\vec{q}_- \times \vec{q}_+)\cdot \vec{k}_+ \; ,
\end{align}
or, in terms of $q_{x,\pm}=q_{T,\pm}\cos\phi_\pm$ and $q_{y,\pm}=q_{T,\pm}\sin\phi_\pm$
\begin{align}
O 
= 2 \; E_+ E_- \; q_{T,+} q_{T,-} \; \sin\Delta \phi_{jj} \; .
\label{eq:phijj_wbf}
\end{align}
where $\Delta \phi_{jj}$ is the signed azimuthal angle difference
\begin{align}
  \Delta \phi_{jj} \equiv \phi_+-\phi_-\; .
\label{eq:signed_wbf}
\end{align}

The main weakness in the observable $O$ is that it depends on the
(usually) poorly determined energies of the initial state partons
$E_\pm$. Rather than relying on the reconstruction of the Higgs
momentum via its decay products to provide this information, we
replace the full observable by
\begin{align}
O \to \Delta \phi_{jj}  \; ,
\label{eq:def_obs_wbf}
\end{align}
which retains the $CP$ sensitivity through the well-defined $\tnaive$
transformation of the full set of observable, matrix element, and
initial state.  The primary difference between the Lorentz-invariant
observable $O$ and $\Delta \phi_{jj}$ is that $O$ is more sensitive to
the magnitude of the tagging jet momenta.  This can be advantageous in
some instances, since the dimension-six operators in the EFT lead to
modifications which grow with momentum transfer. However, the same
effect can be achieved by supplementing $\Delta \phi_{jj}$ with a
virtuality measure such as the transverse momentum of the harder
jet.\bigskip

\begin{figure}[t]
  \centering
  \includegraphics[width=0.49\textwidth]{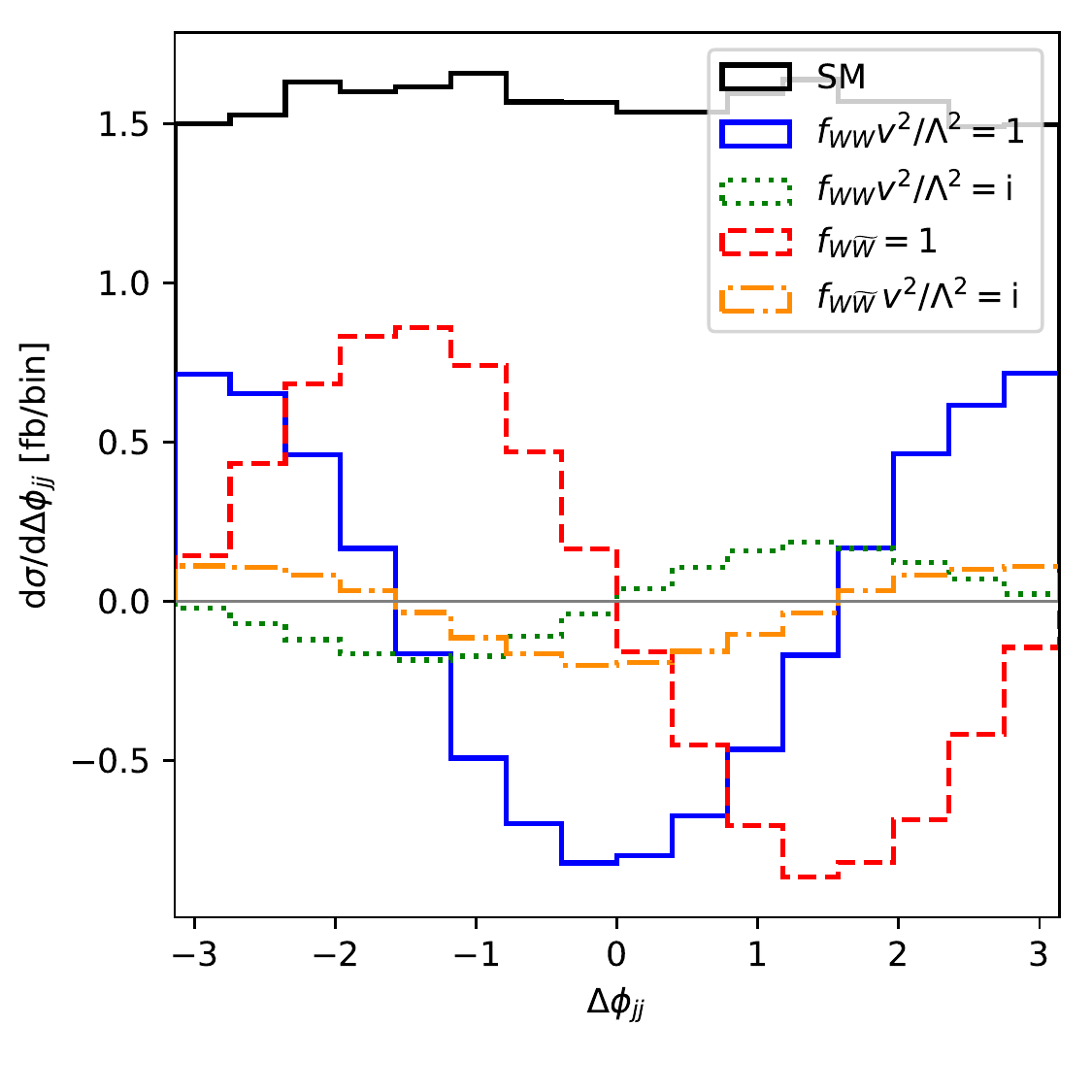}%
  \includegraphics[width=0.49\textwidth]{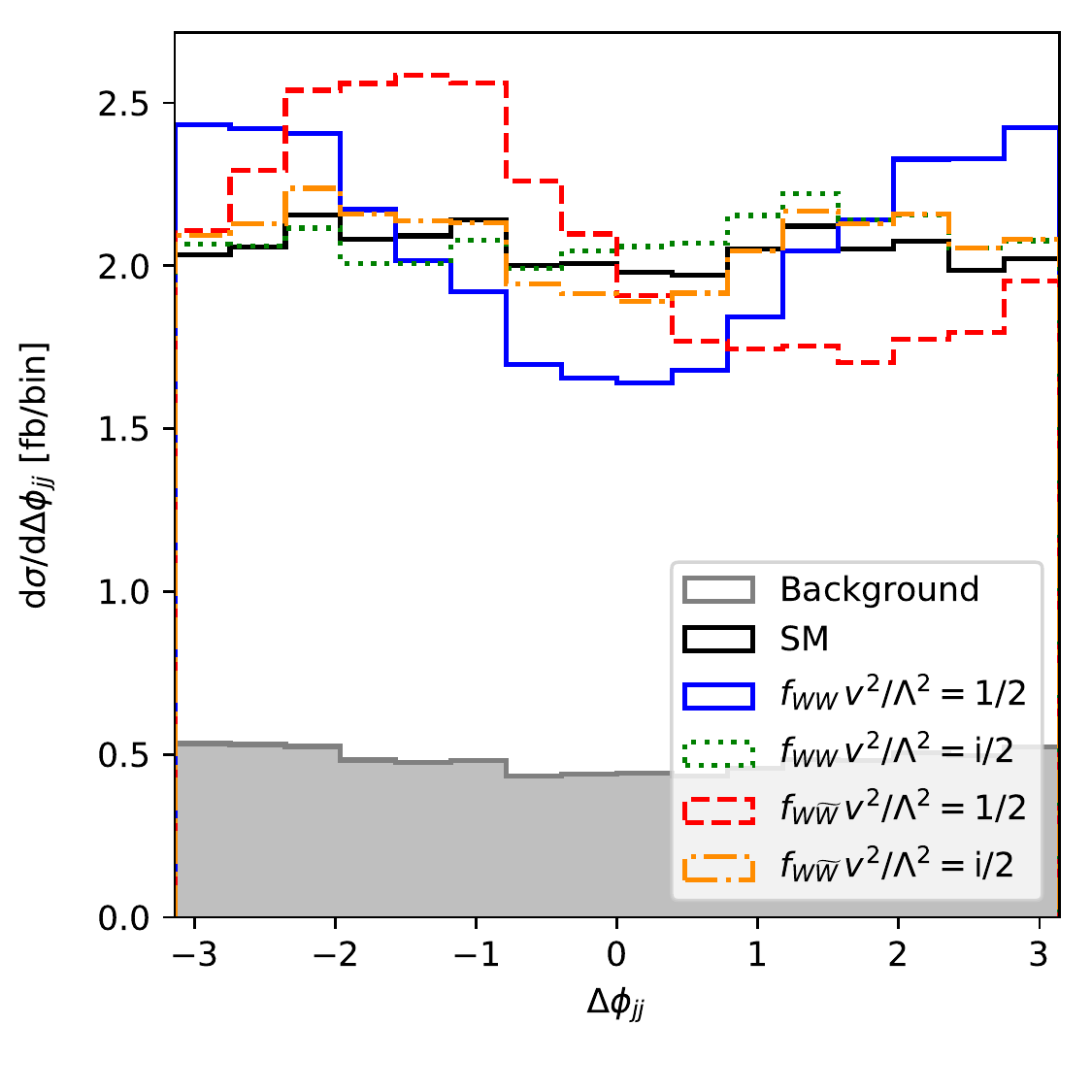}%
  \caption{Distribution of the signed angle $\Delta\phi_{jj}$ in WBF
    Higgs production after the cuts in
    Eqs.\;\eqref{eq:wbf_acceptance_cuts} and
    \eqref{eq:wbf_likelihood_cut} for the Standard Model (black) as well as the
    for the EFT with the indicated Wilson coefficients. In the left
    panel we show the SM signal (black) as well as the
    interference of different dimension-six amplitudes with the SM
    signal (colored).  The right panel shows the full distributions
    including the backgrounds (grey).}
\label{fig:wbf_kinematics}
\end{figure}

We simulate the WBF process with the
\textsc{MadMax}~\cite{madmax2,madmax3} setup of
\textsc{Madgraph}~\cite{madgraph}.  We compute the $\Delta \phi_{jj}$
distribution (with the same event selection as described below)
predicted by the Standard Model as well as for the Standard Model
augmented by representative operators $\ope{WW}$ and
$\ope{W\widetilde{W}}$ defined in Eqs.\;\eqref{eq:op_cpeven} and
\eqref{eq:op_cpodd}.  The resulting distributions are shown in
Fig.~\ref{fig:wbf_kinematics}. As expected, the Standard Model, even
when supplemented by a $CP$-even operator such as $f_{WW}$, results in
a distribution that is symmetric under
$\Delta \phi_{jj} \to - \Delta \phi_{jj}$.  Similar results would be
obtained for the other $CP$-even operators of
Eq.\;\eqref{eq:op_cpeven} such as $\ope{W}$.  In contrast, the
$CP$-odd operator $\ope{W\widetilde{W}}$ leads to a distribution with
a clear preference for $\Delta \phi_{jj} < 0$.

As is evident from Fig.~\ref{fig:wbf_kinematics}, an imaginary Wilson
coefficient $f_{WW}$ also leads to an asymmetry in the
$\Delta \phi_{jj}$ distribution. Clearly, absorptive phases can mimic
the signatures from $CP$-violating scenarios in this non-genuine
$CP$ observable, and thus potentially complicate the interpretation
of such a signature.

\subsection{LHC reach}
\label{sec:wbf_reach}

Based on our simulations, we determine the expected LHC sensitivity to
$\ope{W\widetilde{W}}$ through WBF production followed by the
$H \to \tau \tau$ decay.  The dominant backgrounds are QCD and
electroweak $Zjj$ production followed by the decay $Z \to \tau \tau$,
and Higgs production in gluon fusion with $H \to \tau \tau$.
Our analysis is based on the tagging jet
kinematics~\cite{tagging, wbf_review, wbf_d6}.  We simulate the WBF signal
following Ref.~\cite{madfisher} by generating the process
\begin{align}
  p p \to H \, j j
      \to \tau^+ \tau^- \; j j \, ,
\end{align}
multiplying the rates with the branching ratio for the semi-leptonic
di-tau mode, and assuming the di-tau system to be reconstructed with a
realistic resolution for $m_{\tau\tau}$.  This means that as the
leading detector effect the $m_{\tau \tau}$ distribution is smeared by
a Gaussian~\cite{madmax1,madmax2,madmax3} (with width 17~GeV) for
Higgs production and a double Gaussian (where the dominant component
has a width of 13~GeV) for $Z$ production, as estimated from Fig.~1a
of Ref.~\cite{Aad:2015vsa}. The double Gaussian ensures an accurate
description of the high-mass tail of the $Z$ peak around
$m_{\tau\tau} = m_H$~\cite{johann_thesis}.

Event selection proceeds first with loose cuts
\begin{alignat}{3}
  p_{T,j} &> 20 \ \gev  \qquad & \qquad |\eta_{j}| &< 5.0  \qquad & \qquad 
  \Delta \eta_{jj} &> 2.0  \notag \\ 
  p_{T,\tau} &> 10 \ \gev  \qquad & \qquad |\eta_{\tau}| &< 2.5
\label{eq:wbf_acceptance_cuts}
\end{alignat}
to retain as much phase space information as possible.
One can improve discrimination of the WBF signal from
the electroweak and QCD background processes 
based on their different radiation patterns~\cite{tagging}.
These selections are simulated by applying
central jet veto (CJV) survival probabilities~\cite{wbf_tau},
\begin{align}
  \varepsilon^\text{CJV}_{\text{WBF $H$}} = 0.71 \qqquad
  \varepsilon^\text{CJV}_{\text{EW $Z$}} = 0.48 \qqquad
  \varepsilon^\text{CJV}_{\text{QCD $Z$}} = 0.14 \qqquad
  \varepsilon^\text{CJV}_{\text{GF $H$}} = 0.14 \; .
\end{align}
Provided the hard phase space does not include any jets beyond the two
tagging jets, the results are not expected to first approximation to
be sensitive to details of the central jet veto.  For simplicity, we
assume the reconstruction and identification of the leptonic $\tau$ to
be fully efficient and assume a constant overall efficiency of 0.6 for
the hadronic tau. These efficiencies do not affect the
signal-to-background ratio. As a second way to suppress backgrounds,
we apply a likelihood-based event selection~\cite{madfisher}
\begin{align}
  \frac{\Delta \sigma_\text{SM WBF}(\boldx)}{\Delta \sigma_\text{backgrounds} (\boldx)} > 1 \,,
\label{eq:wbf_likelihood_cut}
\end{align}
retaining only phase-space points $\boldx$ with an expected
signal-to-background ratio of at least unity.

For an integrated luminosity of $L =100~\ifb$, after all efficiencies
and the event selection of Eqs.\;\eqref{eq:wbf_acceptance_cuts} and
\eqref{eq:wbf_likelihood_cut}, we expect a WBF Higgs signal of 1349
events in the Standard Model, together with a total expected
background of 388 events. It is worth noting that these numbers are
optimistic and do not include the full suite of detector effects, fake
backgrounds, etc.\bigskip

We analyze how well WBF production can extract information about $CP$
violation in the dimension-six EFT defined in
Eqs.\;\eqref{eq:op_cpeven} and \eqref{eq:op_cpodd}. The model
parameters of interest are given in Eq.\;\eqref{eq:gspace}.  For these
directions in the EFT parameter space we use the the
\textsc{MadFisher} tools~\cite{madfisher} to find the Fisher
information evaluated at the Standard Model after $L = 100~\ifb$ to be
\begin{align}
  I_{ij} = \hspace*{8pt}
 {\footnotesize 
   \begin{blockarray}{rrrrrrrrrrrrrr}
     \hyperref[eq:op_cpeven]{f_{\phi,2} } & \hyperref[eq:op_cpeven]{f_W }
     & \hyperref[eq:op_cpeven]{f_B}
     & \hyperref[eq:op_cpeven]{f_{WW}} & \hyperref[eq:op_cpeven]{f_{BB}}
     & \hyperref[eq:op_cpodd]{\textcolor{red}{f_{W\widetilde{W}}}}
     & \hyperref[eq:op_cpodd]{\textcolor{red}{f_{B\tilde{B}}}}
     & \hyperref[eq:op_cpeven]{\Imag f_{W} } & \hyperref[eq:op_cpeven]{\Imag f_{B}}
     & \hyperref[eq:op_cpeven]{\Imag f_{WW} } & \hyperref[eq:op_cpeven]{\Imag f_{BB}}
     & \hyperref[eq:op_cpodd]{\textcolor{red}{\Imag f_{W\widetilde{W}}}}
     & \hyperref[eq:op_cpodd]{\textcolor{red}{\Imag f_{B\tilde{B}}}} \\
  \begin{block}{(rrrrrrrrrrrrr)r}
  4942 & -968 & -50 & 54 & 2 & \textcolor{red}{-7} & \textcolor{red}{0} & -1 & 0 & 2 & 0 & \textcolor{red}{36} & \textcolor{red}{0} \topstrut & \hspace*{8pt} \hyperref[eq:op_cpeven]{f_{\phi,2} }\\
  -968 & 715 & 35 & -191 & -3 & \textcolor{red}{1} & \textcolor{red}{0} & 0 & 0 & 0 & 0 & \textcolor{red}{-55} & \textcolor{red}{-1}  & \hspace*{8pt} \hyperref[eq:op_cpeven]{f_{W}}\\
  -50 & 35 & 6 & -9 & 0 & \textcolor{red}{0} & \textcolor{red}{0} & 0 & 0 & 0 & 0 & \textcolor{red}{-2} & \textcolor{red}{0} & \hspace*{8pt} \hyperref[eq:op_cpeven]{f_{B}}\\
  54 & -191 & -9 & 321 & 3 & \textcolor{red}{-1} & \textcolor{red}{0} & 0 & 0 & 1 & 0 & \textcolor{red}{72} & \textcolor{red}{1} & \hspace*{8pt} \hyperref[eq:op_cpeven]{f_{WW}}\\
  2 & -3 & 0 & 3 & 0 & \textcolor{red}{0} & \textcolor{red}{0} & 0 & 0 & 0 & 0 & \textcolor{red}{1} & \textcolor{red}{0} & \hspace*{8pt} \hyperref[eq:op_cpeven]{f_{BB}}\\
  \textcolor{red}{-7} & \textcolor{red}{1} & \textcolor{red}{0} & \textcolor{red}{-1} & \textcolor{red}{0} & \textcolor{red}{359} & \textcolor{red}{4} & \textcolor{red}{41} & \textcolor{red}{1} & \textcolor{red}{-81} & \textcolor{red}{-1} & \textcolor{red}{-1} & \textcolor{red}{0} & \hspace*{8pt} \hyperref[eq:op_cpodd]{\textcolor{red}{f_{W\widetilde{W}}}}\\
  \textcolor{red}{0} & \textcolor{red}{0} & \textcolor{red}{0} & \textcolor{red}{0} & \textcolor{red}{0} & \textcolor{red}{4} & \textcolor{red}{0} & \textcolor{red}{0} & \textcolor{red}{0} & \textcolor{red}{-1} & \textcolor{red}{0} & \textcolor{red}{0} & \textcolor{red}{0} & \hspace*{8pt}\hyperref[eq:op_cpodd]{\textcolor{red}{f_{B\tilde{B}}}}\\
  -1 & 0 & 0 & 0 & 0 & \textcolor{red}{41} & \textcolor{red}{0} & 6 & 0 & -12 & 0 & \textcolor{red}{0} & \textcolor{red}{0} & \hspace*{8pt} \hyperref[eq:op_cpeven]{\Imag f_{W}}\\
  0 & 0 & 0 & 0 & 0 & \textcolor{red}{1} & \textcolor{red}{0} & 0 & 0 & 0 & 0 & \textcolor{red}{0} & \textcolor{red}{0} & \hspace*{8pt} \hyperref[eq:op_cpeven]{\Imag f_{B}}\\
  2 & 0 & 0 & 1 & 0 & \textcolor{red}{-81} & \textcolor{red}{-1} & -12 & 0 & 23 & 0 & \textcolor{red}{0} & \textcolor{red}{0} & \hspace*{8pt} \hyperref[eq:op_cpeven]{\Imag f_{WW} }\\
  0 & 0 & 0 & 0 & 0 & \textcolor{red}{-1} & \textcolor{red}{0} & 0 & 0 & 0 & 0 & \textcolor{red}{0} & \textcolor{red}{0} & \hspace*{8pt} \hyperref[eq:op_cpeven]{\Imag f_{BB}}\\
  \textcolor{red}{36} & \textcolor{red}{-55} & \textcolor{red}{-2} & \textcolor{red}{72} & \textcolor{red}{1} & \textcolor{red}{-1} & \textcolor{red}{0} & \textcolor{red}{0} & \textcolor{red}{0} & \textcolor{red}{0} & \textcolor{red}{0} & \textcolor{red}{21} & \textcolor{red}{0} & \hspace*{8pt} \hyperref[eq:op_cpodd]{\textcolor{red}{\Imag f_{W\widetilde{W}}}}\\
  \textcolor{red}{0} & \textcolor{red}{-1} & \textcolor{red}{0} & \textcolor{red}{1} & \textcolor{red}{0} & \textcolor{red}{0} & \textcolor{red}{0} & \textcolor{red}{0} & \textcolor{red}{0} & \textcolor{red}{0} & \textcolor{red}{0} & \textcolor{red}{0} & \textcolor{red}{0} \botstrut & \hspace*{8pt} \hyperref[eq:op_cpodd]{\textcolor{red}{\Imag f_{B\tilde{B}}}}\\
\end{block}
\end{blockarray}
}
\,,
\end{align}
where red entries correspond to the $CP$-odd coefficients and we
explicitly label the rows and columns with the corresponding Wilson
coefficients.\bigskip

\begin{figure}[t]
  \includegraphics[width=0.49\textwidth]{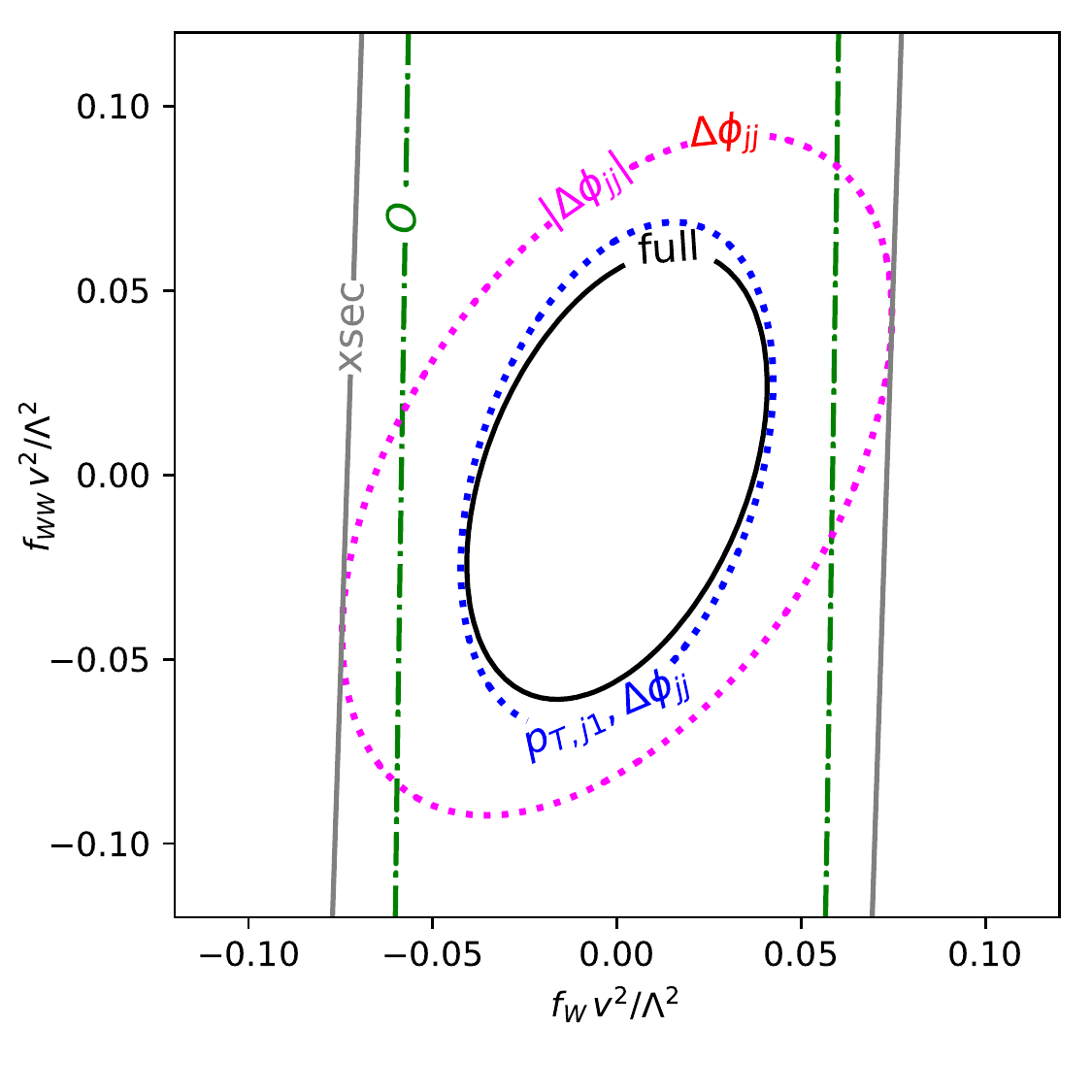}%
  \includegraphics[width=0.49\textwidth]{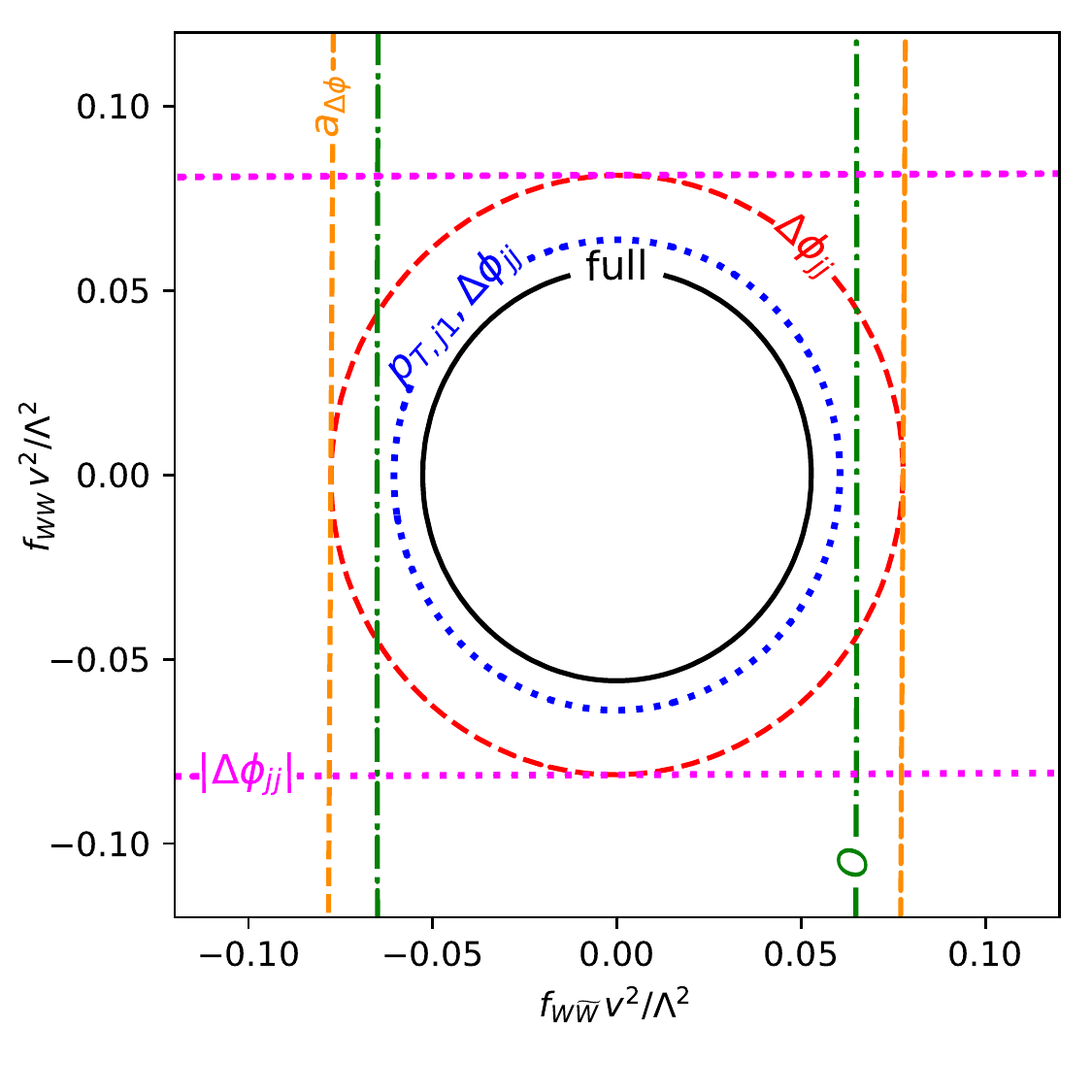}\\%
  \includegraphics[width=0.49\textwidth]{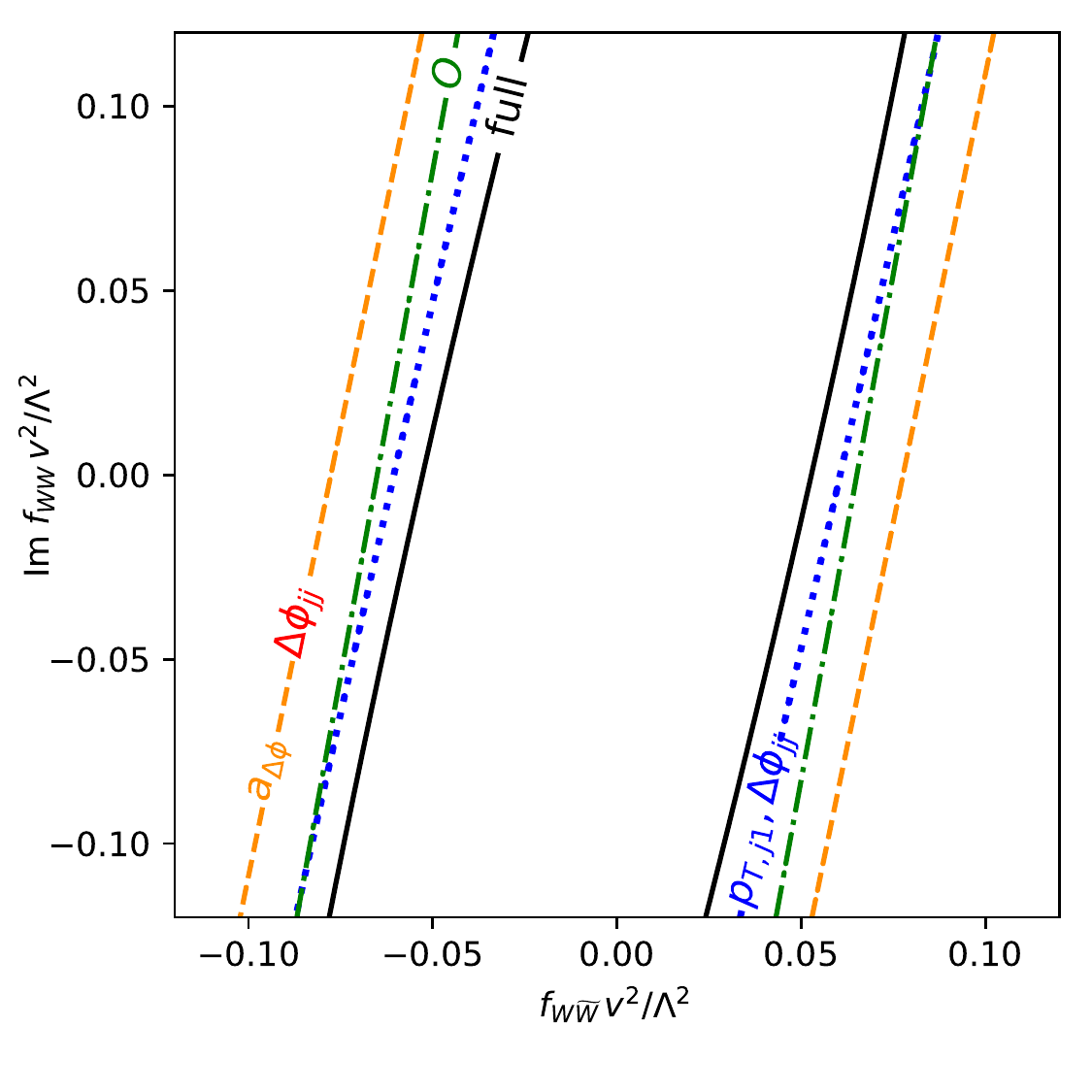}%
  \includegraphics[width=0.49\textwidth]{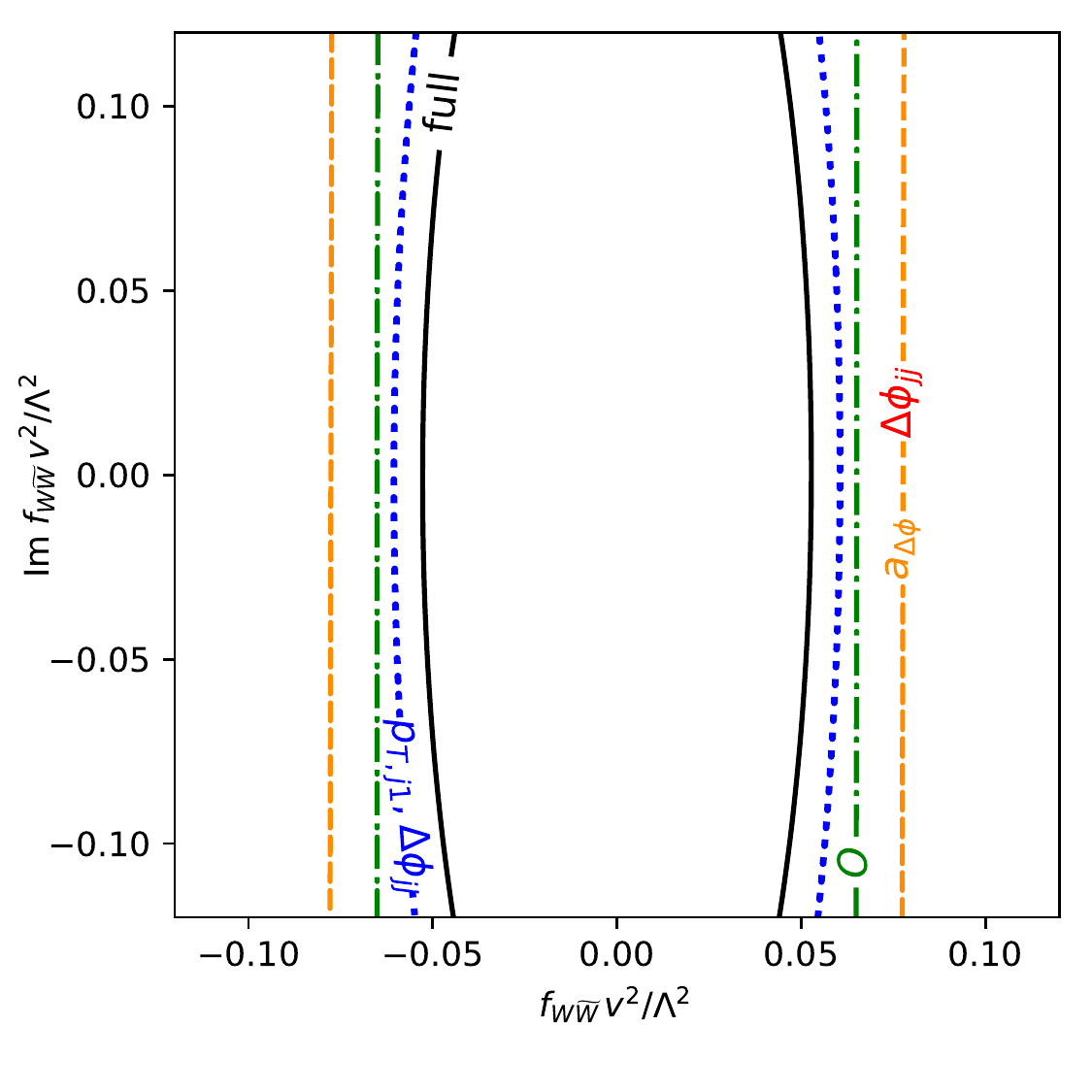}%
  \caption{Optimal $1\sigma$ contours for WBF Higgs production with
    $H \to \tau \tau$ (solid black).  Also shown are the results based
    on different subsets of the $\Delta \phi_{jj}$ distribution,
    including its absolute value (purple), its asymmetry (orange), its
    full distribution (red), its combination with the leading jet
    $p_T$ distribution (blue); as well as the observable $O$ as
    defined in Eq.\;\eqref{eq:phijj_wbf} (green). In grey we show
    bounds based on a simple rate measurement. In each panel, the
    parameters not shown are set to zero.}
  \label{fig:wbf_contours}
\end{figure}

Fig.~\ref{fig:wbf_contours} shows the corresponding optimal error
contours for representative pairs of Wilson coefficients, with those
not shown on the axes set to zero, assuming that the data follows the
Standard Model expectation.  We assume the cuts of
Eqs.\;\eqref{eq:wbf_acceptance_cuts} and \eqref{eq:wbf_likelihood_cut}
and an integrated luminosity of $100~\ifb$.  Since these
two-dimensional combinations of Wilson coefficients may not correspond
to realistic UV scenarios, the projections should be interpreted with
care.

In addition to the full phase-space information, which obviously
results in the best reach, we show the expected constraints from
observables based on subsets of the information contained in the
$\Delta \phi_{jj}$ distribution.  First, we find that all of the
observables are sensitive to various $CP$-even operators. Second, the
signed $\Delta \phi_{jj}$ distribution contains approximately as much
information about $\ope{WW}$ as about $\ope{W\widetilde{W}}$.  In
contrast, the distribution of its absolute value $|\Delta \phi_{jj}|$
is only sensitive to the $CP$-even operators.  In
Fig.~\ref{fig:wbf_contours} we confirm that in the top-left panel the
full $\Delta \phi_{jj}$ results are identical to those from the
absolute value $|\Delta \phi_{jj}|$, while in the two bottom panels
with with their imaginary parts they are identical to those from the
asymmetry
\begin{align}
    a_{\Delta \phi_{jj}} \equiv \frac{\mathrm{d} \sigma (\Delta \phi_{jj})  -  \mathrm{d} \sigma (-\Delta \phi_{jj}) }
{ \mathrm{d} \sigma  (\Delta \phi_{jj}) + \mathrm{d} \sigma  (-\Delta \phi_{jj}) } \,.
\label{eq:def_asymm_wbf}
\end{align}
By definition, this asymmetry is not sensitive to $CP$-even
modification with a real Wilson coefficient. This confirms that any
asymmetry in $\Delta \phi_{jj}$ is a clear indicator of $CP$ violation
as long as we neglect absorptive phases.

The observable $O$ is insensitive to the $CP$-even operator
$\ope{WW}$, because the information in the absolute value
$|\Delta \phi|$ is washed out by the residual momentum dependence.
The same momentum dependence, on the other hand, results in a slightly
enhanced reach for $CP$-violating physics compared to
$a_{\Delta \phi_{jj}}$. This is consistent with the observation that
supplementing $\Delta \phi_{jj}$ with the leading $p_{T,j}$ and
analyzing their joint distribution also significantly improves the
reach.  This enhancement is not per se related to $CP$ violation, but
rather reflects the well known fact that dimension-six operators lead
to effects which are enhanced at higher momentum transfer. These two
important distributions cover the majority of the information after
the selection cuts in Eqs.\;\eqref{eq:wbf_acceptance_cuts} and
\eqref{eq:wbf_likelihood_cut}, with only modest improvements obtained
by including additional phase-space information.

The case of absorptive physics, represented by a complex phase in the Wilson
coefficient, is shown in the lower two panels of Fig.~\ref{fig:wbf_contours}.
While there is some sensitivity to the imaginary parts of $f_{WW}$ and
$f_{W\widetilde{W}}$, the reach is typically much weaker for their
imaginary parts than for the real parts.
Crucially, the lower left panel of Fig.~\ref{fig:wbf_contours}
demonstrates that once we allow for such an absorptive phase, an
almost blind direction in parameter space arises: none of the
observables can unequivocally prove $CP$ violation.\bigskip

To summarize, $CP$-violating scenarios can lead to large asymmetries
in the signed $\Delta \phi_{jj}$ distribution in WBF, giving an
impressive new physics reach of the LHC in these signatures. But this
genuine $\tnaive$-odd observable can only be interpreted as a sign of
$CP$ violation under the additional assumption that re-scattering is
negligible. As a side remark, an essentially equivalent measurement is 
possible for Higgs plus two jets production in gluon fusion, testing the
$CP$ nature of the effective Higgs interaction with gluons~\cite{phi_jj}.

\section{$ZH$ production}
\label{sec:zh}

At the amplitude level and assuming custodial symmetry, the $ZH$
signature is sensitive to the same EFT vertices as WBF
production~\cite{tao1}.  However, its $q \bar q$ initial state is
$CP$-even in the center-of-mass frame and at leading order in
QCD. Following Sec.~\ref{sec:intro_cp_lhc} this implies that one can
construct a genuine $CP$-odd observable~\cite{tao2}.  It thus eschews the need for
additional theory assumptions concerning absorptive phases in the
Wilson coefficients.

We focus on the case with a leptonic $Z$ decay and Higgs decaying into
bottom quarks ,
\begin{align}
q \; \bar{q}  \to Z H \to \ell^- \ell^+ \; b \bar{b}\; ,
\end{align}
which allows us to reconstruct the final state with great precision
including the electric charges of the two leptons, which opens the
door to $C$-sensitive observables.  The specific Higgs decay
$H \to b\bar{b}$ has a large branching ratio, but will not play an
important role in our analysis aside from providing information about
the initial state momenta.

\subsection{CP observables}
\label{sec:zh_obs}

Once again, the lack of access to spins of any of the participants
implies that all realistic observables are constructed from 4-momenta.
Following Sec.~\ref{sec:intro_cp_higgs}, they have the same
transformation properties under $P$ and $\tnaive$, so a $CP$-odd
observable is either $\tnaive$-odd, $P$-odd, and $C$-even, or it is
$\tnaive$-even, $P$-even and $C$-odd.  There are two types of
$CP$ observables distinguished by their transformation under
$\tnaive$:
\begin{enumerate}
\item $CP$-odd and $\tnaive$-odd: as discussed in
  Sec.~\ref{sec:intro_cp_lhc}, there is one $P$-odd, $C$-even
  observable based on the four independent 4-momenta,
\begin{align}
O_1 = \epsilon_{\mu\nu\rho\sigma} \; k_1^\mu k_2^\nu q_{\ell^+}^\rho q_{\ell^-}^\sigma \;
      \text{sign} ( (k_1-k_2) \cdot (q_1-q_2) ) \; ,
\label{eq:p_odd_zh0}
\end{align}
  where $k_{1,2}$ are the initial parton momenta and
  $q_{{\ell^+},{\ell^-}}$ are the outgoing $\ell^+$ and $\ell^-$
  momenta.  As before, the sign ensures that the observable is
  independent of the parton momenta assignment and $C$-even. As in
  Eq.\;\eqref{eq:p_odd_wbf3} and Eq.\;\eqref{eq:phijj_wbf}, $O_1$ can
  be related to the azimuthal angle, for which a sign imposes an
  ordering according to the lepton momentum in the center-of-mass
  frame.
\begin{align}
O_1 \to  \Delta \phi_{\ell\ell} \equiv ( \phi_{\ell^+} - \phi_{\ell^-} ) \; \text{sign}(q_{z,\ell^+}-q_{z,\ell^-})_\text{cm}\; .
\label{eq:p_odd_zh1}
\end{align}

\item $CP$-odd and $\tnaive$-even: the two $C$-odd observables are
  constructed from scalar products between a $C$-even and a $C$-odd
  4-vector.  The $C$-eigenstate 4-vectors are differences of the
  4-momenta,
\begin{align}
  k_\pm \equiv k_1 \pm k_2 \stackrel{C}{\longrightarrow} \pm k_\pm \, , \qqquad 
  q_\pm \equiv q_{\ell^+} \pm q_{\ell^-} \stackrel{C}{\longrightarrow} \pm q_\pm \; .
\end{align}
  Because $(k_+ \cdot k_-) = (q_+ \cdot q_-) = 0$ for massless
  fermions, there are two $C$-odd scalar products ($q_- \cdot k_+$)
  and ($k_- \cdot q_+$), and the remaining four scalar products are
  $C$-even.  The first $C$-odd scalar product maps on to the energy
  difference between the leptons
\begin{align}
O_2 \equiv q_- \cdot k_+  = \sqrt{s} \; (E_{\ell^+}-E_{\ell^-})  \to  (E_{\ell^+}-E_{\ell^-}) \equiv \Delta E_{\ell \ell} \; ,
\label{eq:p_odd_zh2}
\end{align}
  where in the center-of-mass frame $k_{1,2} = (E,0,0,\pm E)$,
  $q_{\ell^\pm} =
  (E_{\ell^\pm},\vec{q}_{T,{\ell^\pm}},q_{z,{\ell^\pm}})$, and $s =
  4E^2$.  The observable ($k_- \cdot q_+$) is a challenge at the LHC,
  because there is no practical way to identify the initial state
  quarks and anti-quarks on an event-by-event basis.  However, the
  $C$-odd combination
\begin{align}
O_3 
& \equiv (q_- \cdot k_+) (q_+ \cdot k_+) - (q_+ \cdot k_-) (q_- \cdot k_-) \notag \\
& = s \, (q_{T,{\ell^+}}-q_{T,{\ell^-}})  (q_{T,{\ell^+}}+q_{T,{\ell^-}})
\to (q_{T,{\ell^+}}-q_{T,{\ell^-}}) \equiv \Delta p_{T,\ell\ell}
\label{eq:p_odd_zh3}
\end{align}
  accesses its information, while only depending on the transverse
  momentum difference of the leptons~\cite{tao2} and the center of
  mass energy, which can be determined once the Higgs momentum is
  reconstructed from its decay products.
\end{enumerate}

Following the discussion in Sec.~\ref{sec:intro_cp_naive}, only the
$\tnaive$-odd observables $O_{1}$ or equivalently $\Delta \phi_{\ell
  \ell}$ probe the $CP$ nature of the Higgs-gauge sector. We
illustrate this in the left panel of Fig.~\ref{fig:zh_kinematics},
which shows the distribution of $\Delta \phi_{\ell \ell}$ for the $ZH$
signal in the SM and with different choices of Wilson coefficients in
the EFT. As expected, the $CP$-odd operator $\ope{W\widetilde{W}}$
induces an asymmetry under $\Delta \phi_{\ell \ell} \to -\Delta
\phi_{\ell \ell}$. Unlike in WBF, this genuine signature of $CP$
violation cannot be generated from an absorptive phase in $CP$-even
physics.

As discussed in Sec.~\ref{sec:intro_cp_naive}, the $\tnaive$-even
observables $O_{2}$ and $O_{3}$ or equivalently $\Delta E_{\ell \ell}$
and $\Delta p_{T,\ell\ell}$ will have a non-zero expectation value
only in the presence of $CP$ violation and re-scattering. The right
panel of Fig.~\ref{fig:zh_kinematics} shows the distribution of
$\Delta E_{\ell \ell}$, demonstrating that an asymmetry in this
observable requires both, $CP$ violation and a source of a complex
phase.

\begin{figure}[t]
  \includegraphics[width=0.49\textwidth]{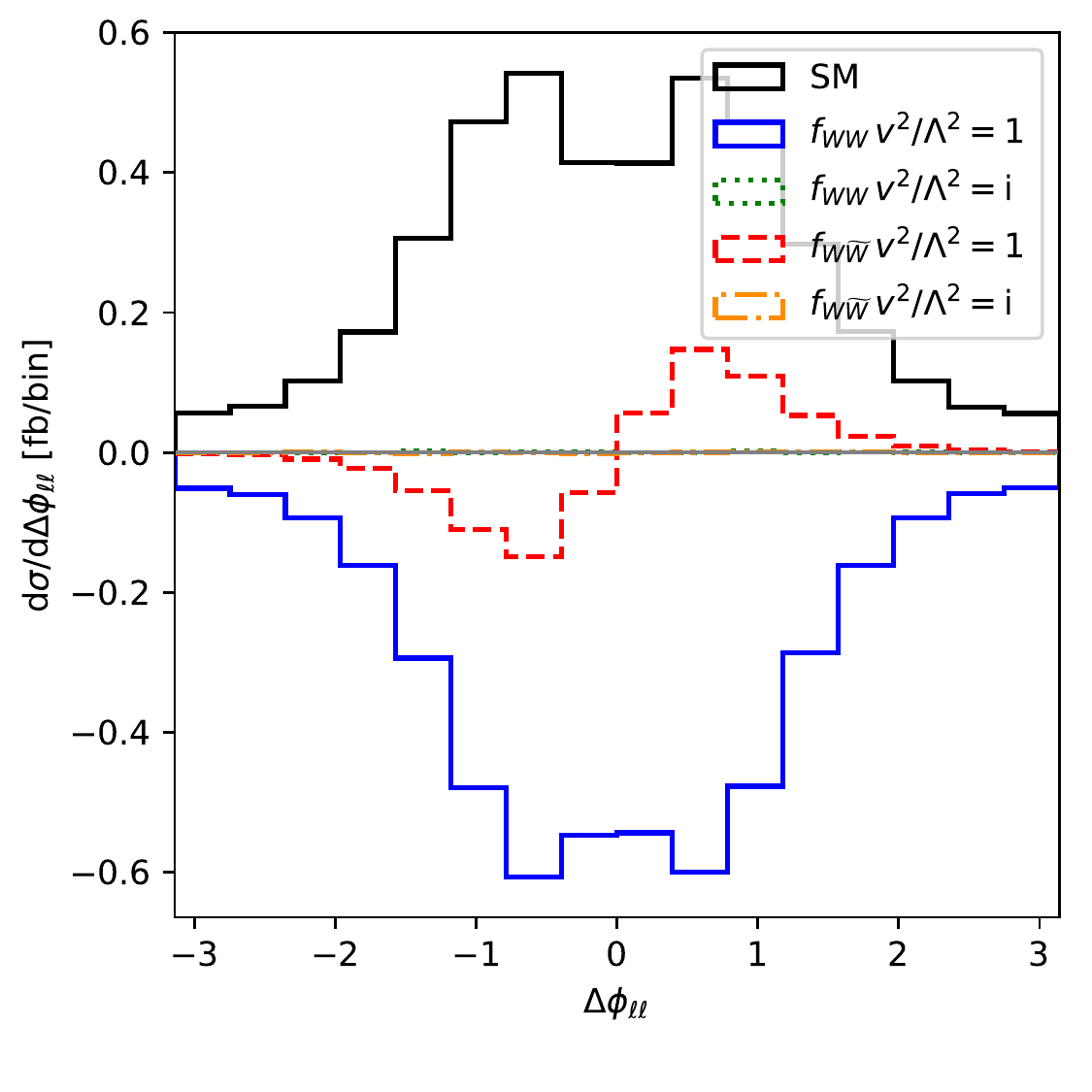}%
  \includegraphics[width=0.49\textwidth]{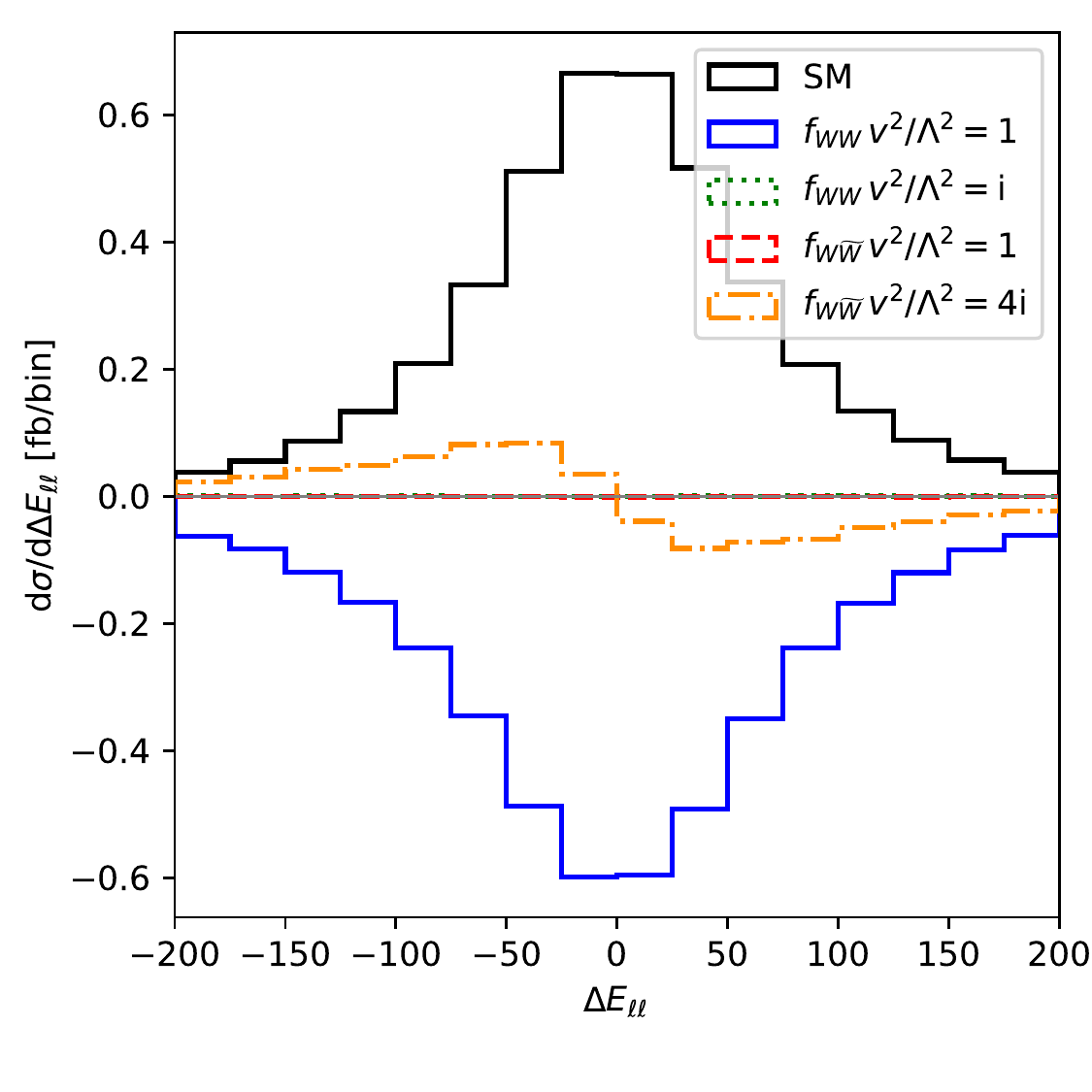}%
  \caption{Distributions of $\Delta\phi_{\ell\ell}$ (left) and
    $\Delta E_{\ell\ell}$ (right) in $ZH$ production after the cuts in
    Eqs.\;\eqref{eq:zh_acceptance_cuts} and
    \eqref{eq:zh_likelihood_cut} for the Standard Model signal (solid
    black), and for the interference between different dimension-six
    amplitudes with the SM signal (colored).}
  \label{fig:zh_kinematics}
\end{figure}

\subsection{LHC reach}
\label{sec:zh_reach}

The signature consists of two $b$-tagged jets and two opposite-sign,
same-flavor leptons.  We simulate it as in Sec.~\ref{sec:wbf}, with
the $b$-jet momenta smeared appropriately for the reconstruction in
the $H \to b \bar{b}$ decay mode with a Gaussian with width
$\sigma_{bb} = 12.5$~GeV~\cite{madmax3}.  The basic acceptance cuts
\begin{alignat}{4}
  p_{T,b} &> 20 \ \gev  \quad & \quad |\eta_b| &< 2.5  \quad & \quad 
  100~\gev &< m_{bb} < 150~\gev \quad & \quad  \Delta R_{bb} &> 0.4 \notag \\ 
  p_{T,\ell} &> 10 \ \gev  \quad & \quad |\eta_{\ell}| &< 2.5 \quad & \quad 
  86~\gev &< m_{\ell \ell} < 96~\gev  \quad & \quad  \Delta R_{\ell\ell}, \Delta R_{\ell b} &> 0.4
\label{eq:zh_acceptance_cuts}
\end{alignat}
include a narrow invariant mass window for the two leptons to
effectively reject background processes without an on-shell $Z \to
\ell \ell$ decay.  After the leptonic invariant mass cut, the main
background is the irreducible $b\bar{b} Z_\ell$ production, where the
two $b$-jets are produced as hadronic radiation.  The acceptance cuts
of Eq.\;\eqref{eq:zh_acceptance_cuts} reduce its rate to 629~fb
(before $b$-tagging), to be compared to the SM $ZH$ signal rate of
14~fb.

We require two $b$-tags. This helps with fake backgrounds (as
explained below), but differs with regard to some of the current
experimental strategies grappling with limited statistics --- a
challenge that is much less of a concern with $100~\ifb$. We assume a
double $b$-tagging rate for the signal and primary background of
$0.7^2$.  Through mis-tagging, the fake QCD background $g g \to c
\bar{c} Z_\ell$ will also contribute. Its rate after the acceptance
cuts is 423~fb, and as long as the rate to mis-tag a charm as a $b$
remains below $20\%$, it is small enough to be ignored.  There is also
a contribution from mis-tagged light-flavor jets.  Starting from a $jj
Z_\ell$ rate of 17.2~pb after acceptance cuts and applying a mis-tag
probability below 1\% it turns out to be negligible.

Pairs of top quarks lead to a final state $b \bar{b} \ell\ell \nu
\bar{\nu}$ which is primarily distinguished from the signal by the
presence of significant $\slashed{E}_T$.  Supplementing the acceptance
cuts with~\cite{tao2,Khachatryan:2014gga}
\begin{align}
\slashed{E}_T < 20~\gev
\label{eq:zh_met_cut}
\end{align}
results in a rate of 13~fb before $b$-tagging. A multi-variate
analysis of the multi-particle final state will further suppress
it to a level where it does not affect the measurement.\bigskip

After the acceptance cuts, requiring two $b$-tags, and the
$\slashed{E}_T$ cut, the only relevant background is therefore
$b\bar{b} Z_\ell$ production.  As in the WBF analysis, we improve the
signal extraction through the likelihood-based event
selection~\cite{madfisher},
\begin{align}
    \frac{\Delta \sigma_\text{SM $ZH$}(\boldx)}{\Delta \sigma_\text{backgrounds} (\boldx)} > 0.1 \; .
  \label{eq:zh_likelihood_cut}
\end{align}
The lower cut-off choice relative to WBF is dictated by the larger
background rates for the $ZH$ case in the relevant phase-space
regions.

With $L =100~\ifb$ of data and after the event selection of
Eqs.\;\eqref{eq:zh_acceptance_cuts}, \eqref{eq:zh_met_cut},
\eqref{eq:zh_likelihood_cut}, and all efficiencies, we expect a $ZH$
signal of 208 events in the Standard Model and a total expected
background of 1035 events. Our idealized treatment of the detector
response and omission of subleading backgrounds mean that these
numbers are certainly optimistic.\bigskip

In the basis of Eq.\;\eqref{eq:gspace}, the Fisher information matrix
evaluated at the Standard Model with an integrated luminosity of
$L=100~\ifb$ is
\begin{align} 
  I_{ij} =\hspace*{8pt}
  {\footnotesize 
    \begin{blockarray}{rrrrrrrrrrrrrr}
      \hyperref[eq:op_cpeven]{f_{\phi,2} } & \hyperref[eq:op_cpeven]{f_W }
      & \hyperref[eq:op_cpeven]{f_B}
      & \hyperref[eq:op_cpeven]{f_{WW}} & \hyperref[eq:op_cpeven]{f_{BB}}
      & \hyperref[eq:op_cpodd]{\textcolor{red}{f_{W\widetilde{W}}}}
      & \hyperref[eq:op_cpodd]{\textcolor{red}{f_{B\tilde{B}}}}
      & \hyperref[eq:op_cpeven]{\Imag f_{W} } & \hyperref[eq:op_cpeven]{\Imag f_{B}}
      & \hyperref[eq:op_cpeven]{\Imag f_{WW} } & \hyperref[eq:op_cpeven]{\Imag f_{BB}}
      & \hyperref[eq:op_cpodd]{\textcolor{red}{\Imag f_{W\widetilde{W}}}}
      & \hyperref[eq:op_cpodd]{\textcolor{red}{\Imag f_{B\tilde{B}}}} \\
      \begin{block}{(rrrrrrrrrrrrr)r}
  4971.9 & 844.2 & 257.5 & 447.0 & 41.6 & \textcolor{red}{0.4} & \textcolor{red}{0.0} & 0.3 & 0.1 & -0.4 & 0.0 & \textcolor{red}{-0.1} & \textcolor{red}{0.0} \topstrut & \hspace*{8pt} \hyperref[eq:op_cpeven]{f_{\phi,2} }\\
  844.2 & 858.5 & 261.8 & 174.2 & 16.2 & \textcolor{red}{0.3} & \textcolor{red}{0.0} & -0.1 & 0.0 & 0.2 & 0.0 & \textcolor{red}{0.3} & \textcolor{red}{0.0}  & \hspace*{8pt} \hyperref[eq:op_cpeven]{f_{W}}\\
  257.5 & 261.8 & 79.8 & 53.1 & 4.9 & \textcolor{red}{0.1} & \textcolor{red}{0.0} & 0.0 & 0.0 & 0.1 & 0.0 & \textcolor{red}{0.1} & \textcolor{red}{0.0}  & \hspace*{8pt} \hyperref[eq:op_cpeven]{f_{B}} \\
  447.0 & 174.2 & 53.1 & 65.8 & 6.1 & \textcolor{red}{0.1} & \textcolor{red}{0.0} & 0.0 & 0.0 & 0.0 & 0.0 & \textcolor{red}{0.0} & \textcolor{red}{0.0} & \hspace*{8pt} \hyperref[eq:op_cpeven]{f_{WW}} \\
  41.6 & 16.2 & 4.9 & 6.1 & 0.6 & \textcolor{red}{0.0} & \textcolor{red}{0.0} & 0.0 & 0.0 & 0.0 & 0.0 & \textcolor{red}{0.0} & \textcolor{red}{0.0}  & \hspace*{8pt} \hyperref[eq:op_cpeven]{f_{BB}}\\
  \textcolor{red}{0.4} & \textcolor{red}{0.3} & \textcolor{red}{0.1} & \textcolor{red}{0.1} & \textcolor{red}{0.0} & \textcolor{red}{5.3} & \textcolor{red}{0.5} & \textcolor{red}{0.0} & \textcolor{red}{0.0} & \textcolor{red}{0.0} & \textcolor{red}{0.0} & \textcolor{red}{0.0} & \textcolor{red}{0.0} & \hspace*{8pt} \hyperref[eq:op_cpodd]{\textcolor{red}{f_{W\widetilde{W}}}}\\
  \textcolor{red}{0.0} & \textcolor{red}{0.0} & \textcolor{red}{0.0} & \textcolor{red}{0.0} & \textcolor{red}{0.0} & \textcolor{red}{0.5} & \textcolor{red}{0.0} & \textcolor{red}{0.0} & \textcolor{red}{0.0} & \textcolor{red}{0.0} & \textcolor{red}{0.0} & \textcolor{red}{0.0} & \textcolor{red}{0.0} & \hspace*{8pt} \hyperref[eq:op_cpodd]{\textcolor{red}{f_{B\widetilde{B}}}}\\
  0.3 & -0.1 & 0.0 & 0.0 & 0.0 & \textcolor{red}{0.0} & \textcolor{red}{0.0} & 5.1 & 1.6 & -7.8 & -0.7 & \textcolor{red}{0.0} & \textcolor{red}{0.0}  & \hspace*{8pt} \hyperref[eq:op_cpeven]{\Imag f_{W}}\\
  0.1 & 0.0 & 0.0 & 0.0 & 0.0 & \textcolor{red}{0.0} & \textcolor{red}{0.0} & 1.6 & 0.5 & -2.4 & -0.2 & \textcolor{red}{0.0} & \textcolor{red}{0.0}  & \hspace*{8pt} \hyperref[eq:op_cpeven]{\Imag f_{B}}\\
  -0.4 & 0.2 & 0.1 & 0.0 & 0.0 & \textcolor{red}{0.0} & \textcolor{red}{0.0} & -7.8 & -2.4 & 12.0 & 1.1 & \textcolor{red}{0.0} & \textcolor{red}{0.0}  & \hspace*{8pt} \hyperref[eq:op_cpeven]{\Imag f_{WW}}\\
  0.0 & 0.0 & 0.0 & 0.0 & 0.0 & \textcolor{red}{0.0} & \textcolor{red}{0.0} & -0.7 & -0.2 & 1.1 & 0.1 & \textcolor{red}{0.0} & \textcolor{red}{0.0}  & \hspace*{8pt} \hyperref[eq:op_cpeven]{\Imag f_{BB}}\\
  \textcolor{red}{-0.1} & \textcolor{red}{0.3} & \textcolor{red}{0.1} & \textcolor{red}{0.0} & \textcolor{red}{0.0} & \textcolor{red}{0.0} & \textcolor{red}{0.0} & \textcolor{red}{0.0} & \textcolor{red}{0.0} & \textcolor{red}{0.0} & \textcolor{red}{0.0} & \textcolor{red}{16.2} & \textcolor{red}{1.5} & \hspace*{8pt} \hyperref[eq:op_cpodd]{\textcolor{red}{\Imag f_{W\widetilde{W}}}}\\
  \textcolor{red}{0.0} & \textcolor{red}{0.0} & \textcolor{red}{0.0} & \textcolor{red}{0.0} & \textcolor{red}{0.0} & \textcolor{red}{0.0} & \textcolor{red}{0.0} & \textcolor{red}{0.0} & \textcolor{red}{0.0} & \textcolor{red}{0.0} & \textcolor{red}{0.0} & \textcolor{red}{1.5} & \textcolor{red}{0.1}  \botstrut & \hspace*{8pt} \hyperref[eq:op_cpodd]{\textcolor{red}{\Imag f_{B\tilde{B}}}}\\
\end{block}
\end{blockarray}
}
\,,
\end{align}
with $CP$-odd components highlighted in red.

This is translated into optimal error contours in
Fig.~\ref{fig:zh_contours}, where each panel shows a pair of Wilson
coefficients, with the remaining ones set to zero. In addition to the
bounds based on the full kinematic information, we also show the
optimal constraints based on individual observables.  Once again, the
best constraints on the $CP$-even operators come from a combination of
angular observables like $\Delta \phi_{\ell \ell}$ and
momentum-sensitive observables like $m_{ZH}$. The $ZH$ production
processes turns out to offer much tighter constraints on $f_W$ than
$f_{WW}$.

The distribution of $\Delta \phi_{\ell\ell}$ is sensitive to both
$CP$-even and $CP$-violating operators. Unsurprisingly, the
information on $CP$-even operators is entirely contained in the
absolute value $|\Delta \phi_{\ell\ell}|$. In contrast, the
differential asymmetry
\begin{align}
    a_{\Delta \phi_{\ell\ell}} \equiv \frac{\mathrm{d} \sigma (\Delta \phi_{\ell\ell})  -  \mathrm{d} \sigma (-\Delta \phi_{\ell\ell}) }
{ \mathrm{d} \sigma  (\Delta \phi_{\ell\ell}) + \mathrm{d} \sigma  (-\Delta \phi_{\ell\ell}) } \, ,
\label{eq:def_asymm_zh}
\end{align}
carries all of the information concerning $CP$ violation. Unlike
$\Delta \phi_{jj}$ in WBF, it is now a genuine $CP$-odd observable, so
this asymmetry is never generated from real or imaginary Wilson
coefficients of $CP$-even operators, and the lower left panel of
Fig.~\ref{fig:zh_contours} does not suffer from blind directions.

As expected from the discussion in Sec.~\ref{sec:intro_cp_lhc}, the
distributions of $\Delta p_{T,\ell\ell}$ and $\Delta E_{\ell\ell}$ can
only exhibit asymmetries if both $CP$ violation and absorptive phases
are present. This leads to these distributions only being sensitive to
the imaginary part of $\ope{W\tilde{W}}$, as is visible in the bottom
right panel of Fig.~\ref{fig:zh_contours}.\bigskip

\begin{figure}
  \centering
  \includegraphics[width=0.49\textwidth]{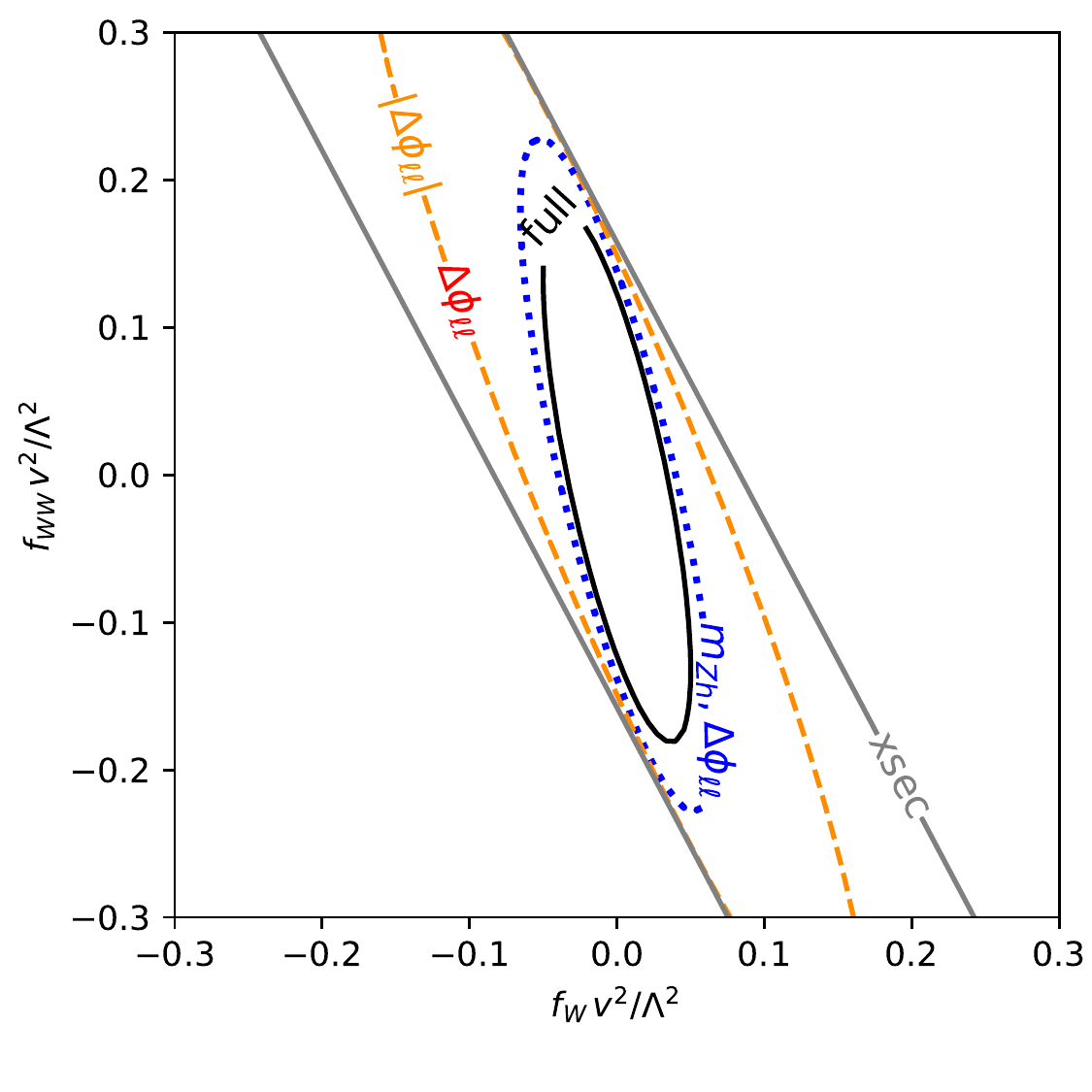}%
  \includegraphics[width=0.49\textwidth]{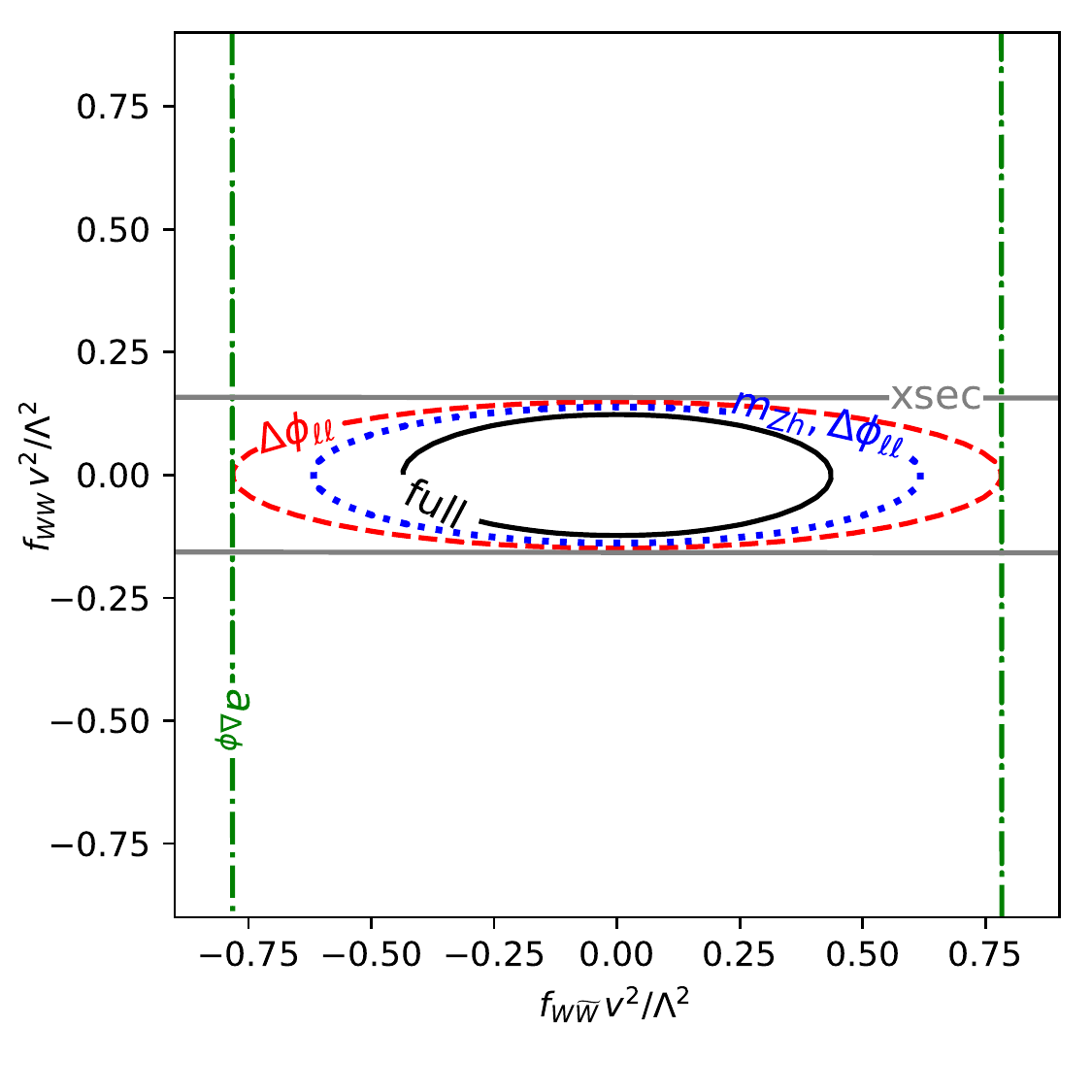}\\%
  \includegraphics[width=0.49\textwidth]{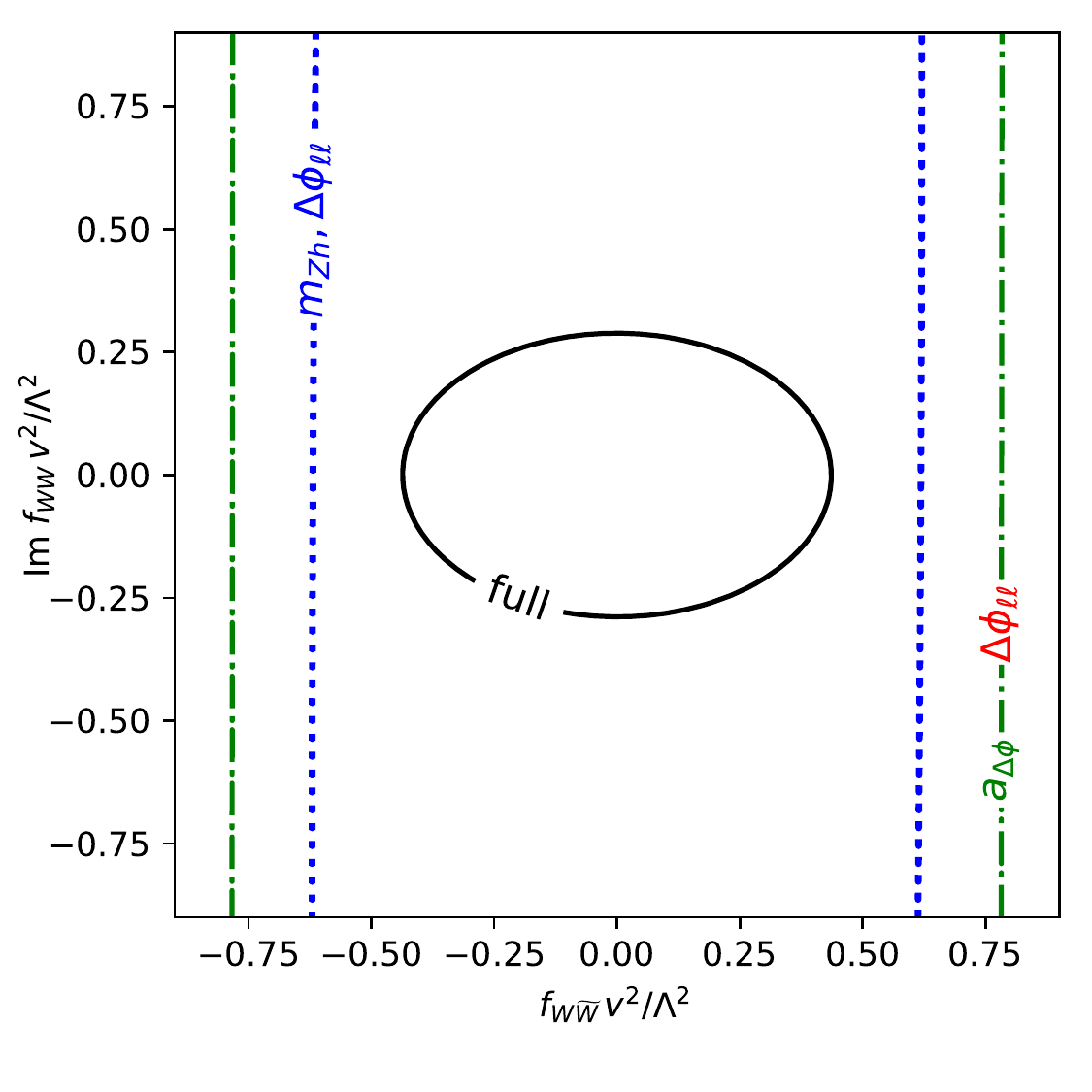}%
  \includegraphics[width=0.49\textwidth]{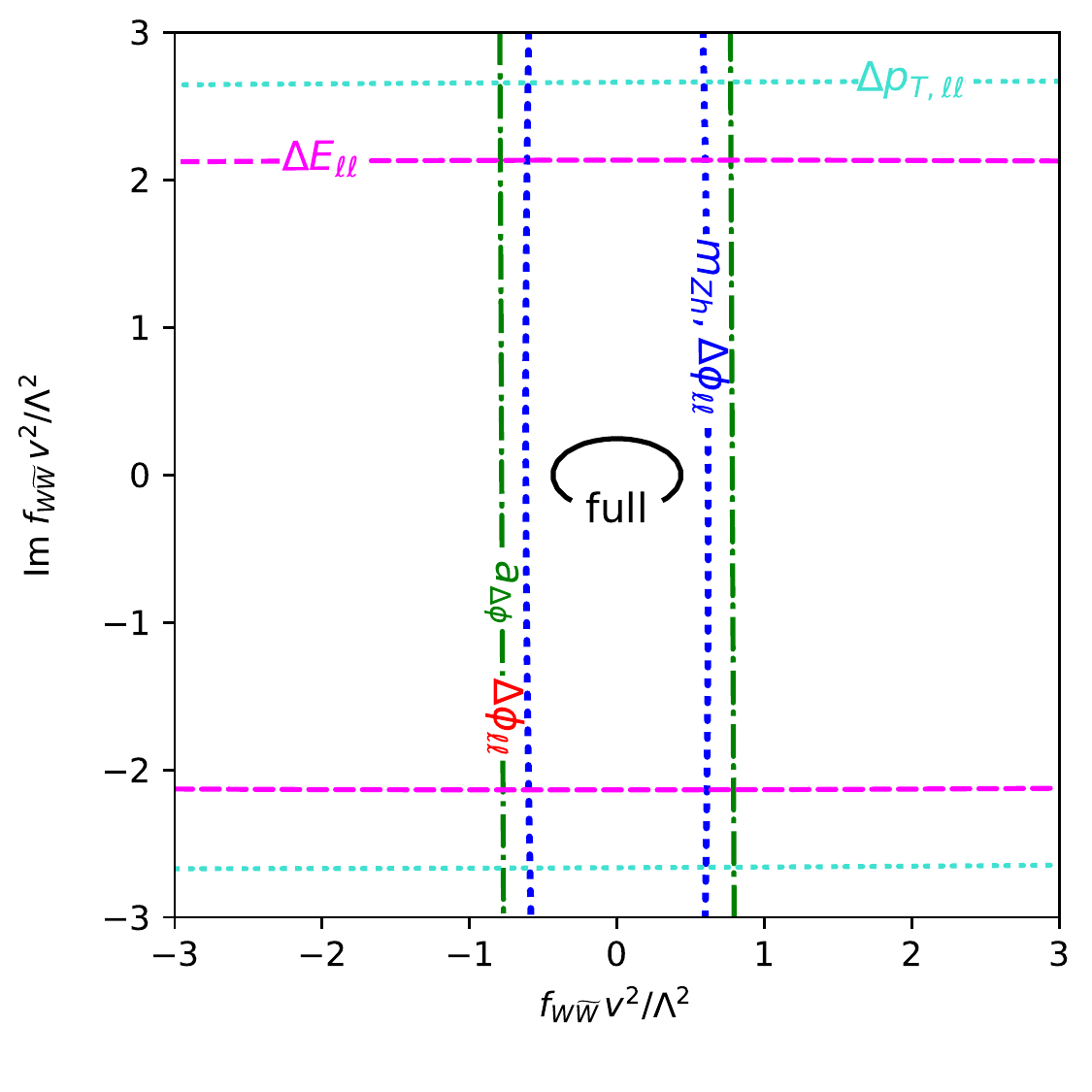}%
  \caption{Optimal $1\sigma$ contours for $ZH$ production (solid
    black).  The colored lines show the reach contained in the
    $\Delta \phi_{\ell\ell}$ distribution, including its absolute
    value (orange), asymmetry (green), full distribution (red),
    combination with the $m_{ZH}$ distribution (blue); based on the
    distribution of $\Delta E_{\ell\ell}$ (purple); for the
    distribution of $\Delta p_{T,\ell\ell}$ (turquoise); and based on
    a simple rate measurement (grey). In each panel, the parameters
    not shown are set to zero.}
  \label{fig:zh_contours}
\end{figure}

Altogether, we find that $ZH$ production with its $CP$-even initial
state provides us with genuine $CP$-odd observables that do not rely
on any further theory assumptions. In particular the signed azimuthal
angle difference $\Delta \phi_{\ell\ell}$ provides a clean probe of
the $CP$ nature of the Higgs-gauge sector. Unfortunately, the small
rate and large backgrounds limit the new physics reach of the LHC in
this channel.

\section{$H \to 4$~leptons}
\label{sec:dec}

The final, classic~\cite{cabibbo,nelson} process we consider is the
Higgs decaying into four leptons, which offers full reconstruction of
the final state, including all of the electric charges.  On the other
hand, in this process the Higgs is almost always on-shell, limiting
the momentum flow through the $HZZ$ vertex. In addition, the fact that
one of the $Z$ bosons is typically on-shell results in one less
independent degree of freedom in the favored region of kinematics.
Nevertheless, the four leptons in the final state can be reconstructed
with exquisite precision, which might help compensate for the smaller
lever arm in energy.

The four leptons are organized into two same-flavor, opposite-sign
pairs, whose momenta are labeled as:
\begin{align}
H \to ~\ell_1^+ (q_{11}) ~\ell_1^- (q_{12}) ~ \ell_2^+ (q_{21}) ~ \ell_2^- (q_{22}) \; .
\end{align}
where $\ell_{1,2} = e,\mu$ are restricted to electrons and muons which
can be reconstructed very precisely.  Even combined with the
relatively featureless gluon-fusion Higgs production mode, this decay
mode has essentially no backgrounds and is largely statistics
limited.

\subsection{CP observables}
\label{sec:dec_obs}

Once again, the lack of spin information dictates that all observables
are constructed from the 4-momenta, and transformation the same way
under both $P$ and $\tnaive$.  Thus, as for $ZH$ production, any
$CP$-odd observable is either $\tnaive$-odd, $C$-even, and $P$-odd and
or $\tnaive$-even, $C$-odd, and $P$-even.  The initial state, at
leading order, is $CP$-symmetric and $\tnaive$-symmetric in the Higgs
rest or center-of-mass frames.  We combine the lepton 4-momenta into
$C$-eigenstates
\begin{align}
q_{1\pm} = q_{11}\pm q_{12} \qqquad 
q_{2\pm} = q_{21}\pm q_{22} \; .
\label{eq:momenta_dec}
\end{align}
Similarly to the discussion in Secs.~\ref{sec:intro_cp_lhc} and \ref{sec:wbf_obs},
there are two classes of observables:
\begin{enumerate}
\item $CP$-odd and $\tnaive$-odd: there is exactly one observable in
  $H\to 4\ell$ decays that is $P$-odd and $C$-even,
\begin{align}
O_a \equiv 
4\epsilon_{\mu\nu\rho\sigma} \; q_{11}^\mu ~q_{12}^\nu ~q_{21}^\rho ~q_{22}^\sigma
=  \epsilon_{\mu\nu\rho\sigma} \; q_{1+}^\mu ~q_{1-}^\nu ~q_{2+}^\rho ~q_{2-}^\sigma \; .
\end{align}
  Unlike in Eq.\;\eqref{eq:p_odd_wbf2}, there is no need for an explicit sign
  factor to compensate for unobservable permutations. It is convenient to
  work in the Higgs rest frame with both $Z$-boson 3-momenta along the
  $z$-axis, implying $q_{i+}=(E_{i},0,0,q_{z,i})$ with $E_1 +E_2=m_H$ and 
  $q_{z,1} + q_{z,2}=0$.  In this frame,
\begin{align}
O_a \rightarrow m_H ~(\vec{q}_{1-} \times \vec{q}_{2-}) \cdot \vec{q}_{1+} \; .
\end{align}
We can relate this to the $Z$-decay plane correlation angle $\Phi$ in
Eq.(2) of Ref.~\cite{hopkins} by introducing
$\vec{n}_i = \vec{q}_{i1}\times \vec{q}_{i2}$ and making use of the
identity
$(\vec{a}\times \vec{b})\times (\vec{a}\times \vec{c}) =
\big[\vec{a}\cdot (\vec{b}\times \vec{c}) \big] \vec{a}$, leading to
\begin{align}
O_a  
= - \frac{16 m_H}{\vec{q}_{1+}^2}  \; \vec{q}_{1+} \cdot \left(\vec{n}_1\times\vec{n}_2 \right) \sim \sin \Phi \; .
\end{align}
  Note that in Ref.~\cite{hopkins} the definition of the angle between
  the two $Z$-decay planes is slightly more complicated. They relate
  the absolute value of $\Phi$ from $\cos \Phi = (\vec{n_1}
  \vec{n}_2)/|\vec{n_1} \vec{n}_2|$ and extract the sign of the angle
  from sign$(\Phi)= \vec{q}_{1+} \cdot (\vec{n}_1\times\vec{n}_2)/(|
  \vec{q}_{1+} \cdot (\vec{n}_1\times\vec{n}_2 ) |)$ .  Since only the
  latter is sensitive to $P$ violation, its information is equivalent
  to $O_a$.
\item $CP$-odd and $\tnaive$-even: as before, we construct two
  scalar-product-based $CP$-odd observables by combining $C$-even and
  $C$-odd 4-vectors: $(q_{2+} \cdot q_{1-})$ and $(q_{1+} \cdot
  q_{2-})$.  In the rest frame of $q_{1+}$ we define
\begin{align}
O_b \equiv - (q_{2+} \cdot q_{1-}) 
    = 2 |\vec{q}_{2+}| |\vec{q}_{11}| \cos\theta_1  \; ,
\end{align}
  with the same angle $\theta_1$ as in 
  Ref.~\cite{hopkins}. Similarly, in the $q_{2+}$ rest frame, we define
\begin{align}
O_c \equiv - (q_{1+} \cdot q_{2-}) 
    = 2 |\vec{q}_{1+}| |\vec{q}_{21}| \cos\theta_2 \; .
\end{align}
\end{enumerate}
The relation between these decay angles the tagging jet correlation in
WBF is well known~\cite{our_nelson}.  Because the effects of
dimension-six operators are enhanced at higher momentum transfer,
selections on the invariant masses $q_{1+}^2$ and $q_{2+}^2$ can
enhance the sensitivity to $CP$-violating operators, even though these
variables themselves are not sensitive to $CP$ violation.

\subsection{LHC reach}
\label{sec:dec_reach}

We simulate the signature 
\begin{align}
pp \to H \to \ell_1^+ \ell_1^- \; \ell_2^+ \ell_2^- 
\end{align}
with two pairs of opposite-sign, same-flavor leptons $\ell_{1,2} = e, \mu$. 
We apply the basic event selection
\begin{align}
  p_{T,\ell} > 10~\gev \qqquad
  |\eta_{\ell}| < 2.5 \qqquad
  120~\gev < m_{4\ell} < 130~\gev \; .
  \label{eq:4l_acceptance_cuts}
\end{align}
After these cuts, there is a small background from continuum $ZZ$
production, which we include with an appropriate smearing of the
$m_{4\ell}$ invariant masses.

As before, we assume an integrated luminosity of $100~\ifb$ and
neglect the detector efficiencies for the four leptons.  For the
Wilson coefficients given in Eq.\;\eqref{eq:gspace}, we find the
following Fisher information matrix evaluated at the Standard Model,
with $CP$-odd components in red:
\begin{align}
  I_{ij} = \hspace*{8pt}
 {\footnotesize 
   \begin{blockarray}{rrrrrrrrrrrrrr}
     \hyperref[eq:op_cpeven]{f_{\phi,2} } & \hyperref[eq:op_cpeven]{f_W }
     & \hyperref[eq:op_cpeven]{f_B}
     & \hyperref[eq:op_cpeven]{f_{WW}} & \hyperref[eq:op_cpeven]{f_{BB}}
     & \hyperref[eq:op_cpodd]{\textcolor{red}{f_{W\widetilde{W}}}}
     & \hyperref[eq:op_cpodd]{\textcolor{red}{f_{B\tilde{B}}}}
     & \hyperref[eq:op_cpeven]{\Imag f_{W} } & \hyperref[eq:op_cpeven]{\Imag f_{B}}
     & \hyperref[eq:op_cpeven]{\Imag f_{WW} } & \hyperref[eq:op_cpeven]{\Imag f_{BB}}
     & \hyperref[eq:op_cpodd]{\textcolor{red}{\Imag f_{W\widetilde{W}}}}
     & \hyperref[eq:op_cpodd]{\textcolor{red}{\Imag f_{B\tilde{B}}}} \\
  \begin{block}{(rrrrrrrrrrrrr)r}
1649.4 & 135.1 & 41.4 & -69.7 & -6.4 & \textcolor{red}{0.3} & \textcolor{red}{0.1} & -1.4 & 0.3 & 2.1 & -0.1 & \textcolor{red}{-0.2} & \textcolor{red}{0.1}  \topstrut & \hspace*{8pt} \hyperref[eq:op_cpeven]{f_{\phi,2} }\\
  135.1 & 11.4 & 3.5 & -6.2 & -0.5 & \textcolor{red}{0.0} & \textcolor{red}{0.0} & -0.1 & 0.0 & 0.2 & 0.0 & \textcolor{red}{0.0} & \textcolor{red}{0.0} & \hspace*{8pt} \hyperref[eq:op_cpeven]{f_{W}}\\
  41.4 & 3.5 & 1.1 & -1.8 & -0.2 & \textcolor{red}{0.0} & \textcolor{red}{0.0} & 0.0 & 0.0 & 0.0 & 0.0 & \textcolor{red}{0.0} & \textcolor{red}{0.0} & \hspace*{8pt} \hyperref[eq:op_cpeven]{f_{B}} \\
  -69.7 & -6.2 & -1.8 & 3.9 & 0.3 & \textcolor{red}{0.0} & \textcolor{red}{0.0} & 0.1 & 0.0 & -0.1 & 0.0 & \textcolor{red}{0.0} & \textcolor{red}{0.0} & \hspace*{8pt} \hyperref[eq:op_cpeven]{f_{WW}} \\
  -6.4 & -0.5 & -0.2 & 0.3 & 0.2 & \textcolor{red}{0.0} & \textcolor{red}{0.0} & 0.0 & 0.0 & 0.0 & 0.0 & \textcolor{red}{0.0} & \textcolor{red}{0.0} & \hspace*{8pt} \hyperref[eq:op_cpeven]{f_{BB}} \\
  \textcolor{red}{0.3} & \textcolor{red}{0.0} & \textcolor{red}{0.0} & \textcolor{red}{0.0} & \textcolor{red}{0.0} & \textcolor{red}{0.5} & \textcolor{red}{-0.1} & \textcolor{red}{0.0} & \textcolor{red}{0.0} & \textcolor{red}{0.0} & \textcolor{red}{0.0} & \textcolor{red}{0.0} & \textcolor{red}{0.0} & \hspace*{8pt} \hyperref[eq:op_cpodd]{\textcolor{red}{f_{W\widetilde{W}}}} \\
  \textcolor{red}{0.1} & \textcolor{red}{0.0} & \textcolor{red}{0.0} & \textcolor{red}{0.0} & \textcolor{red}{0.0} & \textcolor{red}{-0.1} & \textcolor{red}{0.1} & \textcolor{red}{0.0} & \textcolor{red}{0.0} & \textcolor{red}{0.0} & \textcolor{red}{0.0} & \textcolor{red}{0.0} & \textcolor{red}{0.0} & \hspace*{8pt} \hyperref[eq:op_cpodd]{\textcolor{red}{f_{B\widetilde{B}}}} \\
  -1.4 & -0.1 & 0.0 & 0.1 & 0.0 & \textcolor{red}{0.0} & \textcolor{red}{0.0} & 0.0 & 0.0 & -0.1 & 0.0 & \textcolor{red}{0.0} & \textcolor{red}{0.0} & \hspace*{8pt} \hyperref[eq:op_cpeven]{\Imag f_{W}} \\
  0.3 & 0.0 & 0.0 & 0.0 & 0.0 & \textcolor{red}{0.0} & \textcolor{red}{0.0} & 0.0 & 0.0 & 0.1 & 0.0 & \textcolor{red}{0.0} & \textcolor{red}{0.0} & \hspace*{8pt} \hyperref[eq:op_cpeven]{\Imag f_{B}} \\
  2.1 & 0.2 & 0.0 & -0.1 & 0.0 & \textcolor{red}{0.0} & \textcolor{red}{0.0} & -0.1 & 0.1 & 0.3 & 0.0 & \textcolor{red}{0.0} & \textcolor{red}{0.0} & \hspace*{8pt} \hyperref[eq:op_cpeven]{\Imag f_{WW}} \\
  -0.1 & 0.0 & 0.0 & 0.0 & 0.0 & \textcolor{red}{0.0} & \textcolor{red}{0.0} & 0.0 & 0.0 & 0.0 & 0.1 & \textcolor{red}{0.0} & \textcolor{red}{0.0} & \hspace*{8pt} \hyperref[eq:op_cpeven]{\Imag f_{BB}} \\
  \textcolor{red}{-0.2} & \textcolor{red}{0.0} & \textcolor{red}{0.0} & \textcolor{red}{0.0} & \textcolor{red}{0.0} & \textcolor{red}{0.0} & \textcolor{red}{0.0} & \textcolor{red}{0.0} & \textcolor{red}{0.0} & \textcolor{red}{0.0} & \textcolor{red}{0.0} & \textcolor{red}{0.4} & \textcolor{red}{0.0} & \hspace*{8pt} \hyperref[eq:op_cpodd]{\textcolor{red}{\Imag f_{W\widetilde{W}}}} \\
  \textcolor{red}{0.1} & \textcolor{red}{0.0} & \textcolor{red}{0.0} & \textcolor{red}{0.0} & \textcolor{red}{0.0} & \textcolor{red}{0.0} & \textcolor{red}{0.0} & \textcolor{red}{0.0} & \textcolor{red}{0.0} & \textcolor{red}{0.0} & \textcolor{red}{0.0} & \textcolor{red}{0.0} & \textcolor{red}{0.1}  \botstrut & \hspace*{8pt} \hyperref[eq:op_cpodd]{\textcolor{red}{\Imag f_{B\tilde{B}}}} \\
\end{block}
\end{blockarray}
}
\; .
\end{align}

Optimal exclusion limits for representative pairs of Wilson
coefficients are shown in Fig.~\ref{fig:4l_contours}. The sensitivity
is about ten times worse for $H \to 4 \ell$ than for $ZH$ or WBF
production, indicating that the enhanced precision in measuring the
lepton momenta does not overcome the limitations of the restricted
momentum transfer and dominantly constrained kinematics.

\begin{figure}
  \centering
  \includegraphics[width=0.49\textwidth]{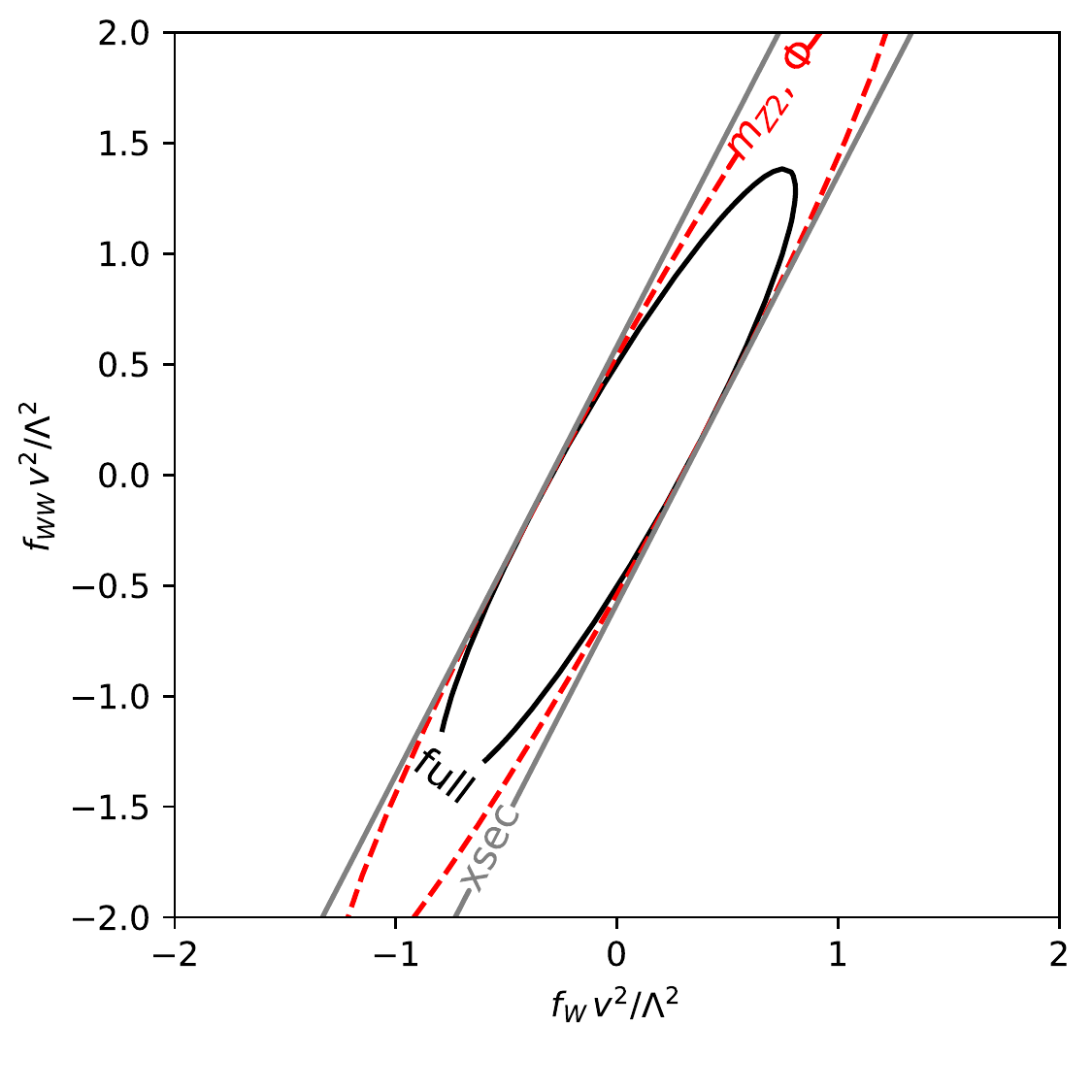}%
  \includegraphics[width=0.49\textwidth]{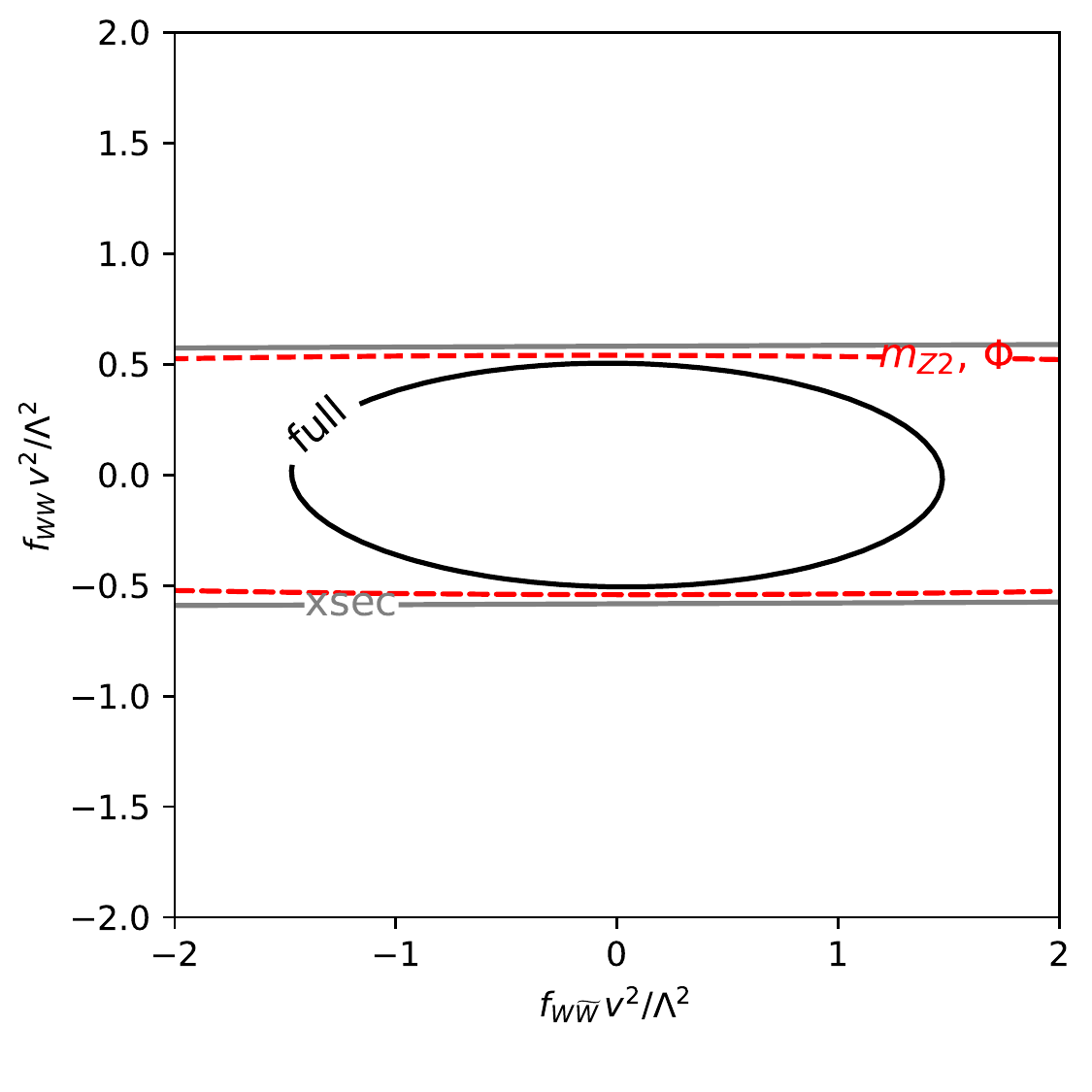}%
  \caption{Optimal $1\sigma$ contours for $gg \to h\to 4 \ell$ (solid
    black).  In grey we show bounds based on a rate measurement. The
    red line shows the contours based on an analysis of the lower of
    the two lepton pair masses $m_{Z_2}$ and $\Phi$, other kinematic
    variables lead to bounds between the red and grey lines. In each
    panel, the parameters not shown are set to zero.}
  \label{fig:4l_contours}
\end{figure}

\section{Comparison and summary}
\label{sec:summary}

The presence of new sources of $CP$ violation in the Higgs sector is
of fundamental importance, and according to some common lore may shed
light on mysteries such as the baryon asymmetry of the Universe.  It
is crucial to establish the symmetry structure through well-defined
observables as an ingredient to a global analysis, for example a
dimension-six effective field theory containing both $CP$-conserving
and $CP$-violating operators.

\begin{figure}
  \centering
  \includegraphics[width=\textwidth]{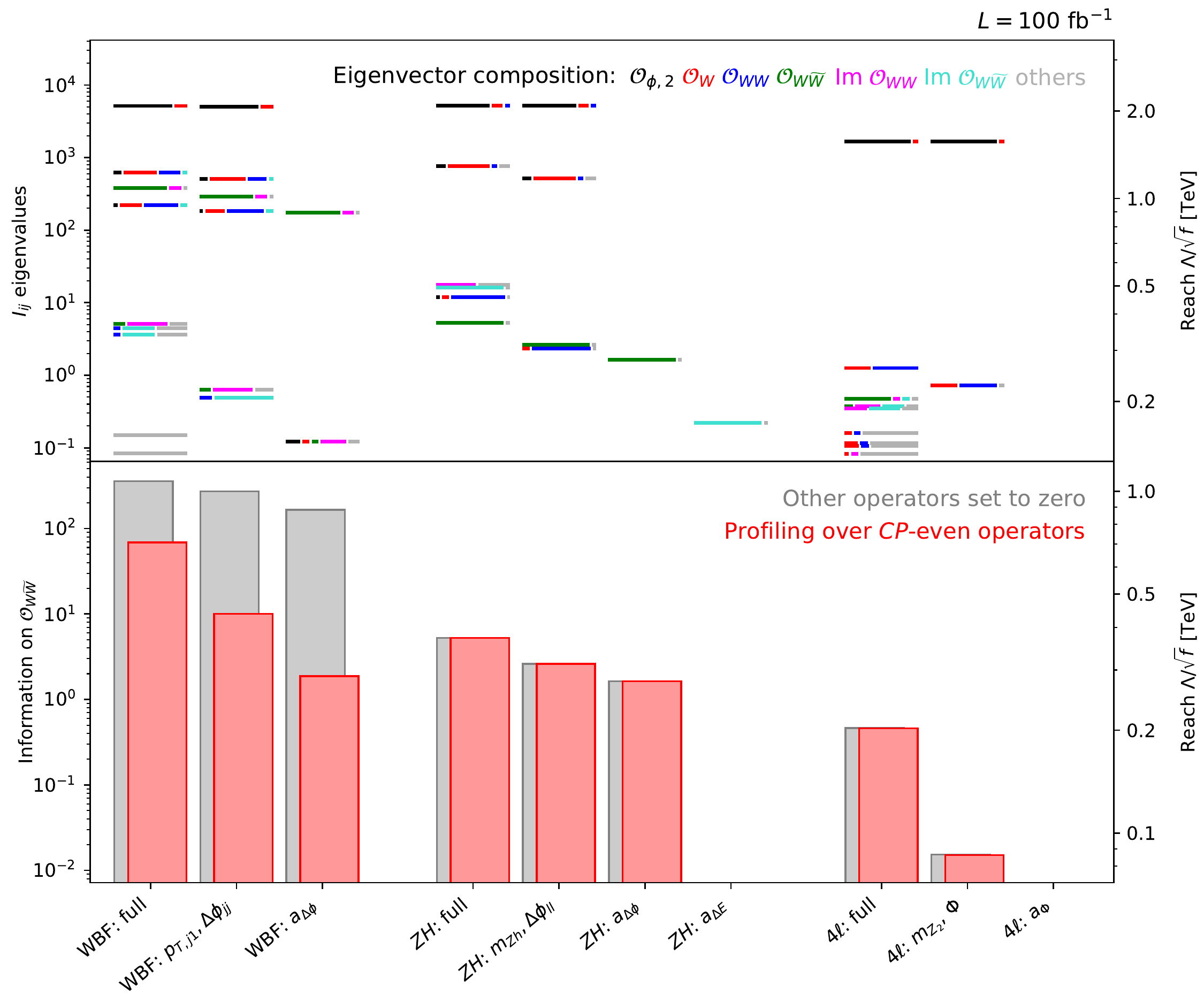}%
  \caption{Comparison of the sensitivity of different channels and
    observables at the LHC with $100~\ifb$. In the top panel we show
    the eigenvalues of the various Fisher information matrices. The
    colors denote the decomposition of the corresponding eigenvectors:
    the length of each segment is proportional to the magnitude of the
    eigenvector component.  In the bottom panel we show the Fisher
    information on the $CP$-violating Wilson coefficient
    $f_{W\widetilde{W}}$. The grey bars show the sensitivity assuming
    that all other considered operators are zero, while the red bars
    profile over arbitrary values of all of the $CP$-even parameters
    (including absorptive parts). In both panels, the right axes
    translate the Fisher information into the corresponding new
    physics reach.}
  \label{fig:comparison}
\end{figure}

We have examined $CP$-sensitive observables in WBF Higgs production,
$ZH$ production, and Higgs decays into four leptons. While the
underlying hard processes, and hence the sensitivity to the $CP$
properties of the Higgs-gauge sector, are essentially identical for
the three processes, the different initial and final state assignments
define distinct signatures:
\begin{enumerate}
\item For WBF the initial state is not a $CP$ eigenstate and one
  cannot measure the charges of initial-state or final-state
  quarks. In this situation we can use the naive time reversal to test
  the underlying $CP$ properties, but only under the assumption of no
  re-scattering effects. On the other hand, the momentum flow through
  the Higgs vertex can be large.
\item In $ZH$ production, the initial state is a $CP$ eigenstate at
  leading order and one can easily identify the lepton charges in the
  final state. We can construct a genuine $CP$-odd observable,
  which directly reflects the $CP$ symmetry of the underlying
  Lagrangian without any assumptions and without any additional
  complex phases. The momentum flow through the Higgs vertex can be
  enhanced by kinematic cuts.
\item Finally, for $H \to e^+ e^- \mu^+ \mu^-$, one has full control
  over the kinematics of the process, allowing for a straightforward
  construction of $CP$-sensitive observables. However, the momentum
  flow through the relevant Higgs vertex is restricted by the Higgs
  mass and one of the $Z$-bosons is on-shell, limiting the kinematic
  coverage of the process.
\end{enumerate}

In a next step, we have analyzed the new physics reach of the
processes and observables in terms of thirteen Wilson coefficients.
By calculating the Fisher information in the different signatures, we
determined the optimal possible exclusion limits at the LHC, including
through any multivariate analysis and taking into account all
correlations between different operators.

The results of this comprehensive comparison are summarized in
Fig.~\ref{fig:comparison}. We compare the optimal sensitivity of the
three analyzed channels when either performing a fully multivariate
analysis or a histogram-based analysis of either one or a combination
of two kinematic distributions, assuming an integrated luminosity of
$100~\ifb$. In the top panel we show the eigenvalues and eigenvectors
of the Fisher information matrices. The colors denote the
decomposition of the corresponding eigenvectors, defining the
direction of a given eigenvector in model parameter space.  The right
axis translates the corresponding Fisher information into the new
physics reach along this direction in model parameter space.

In general, the $CP$-even operator $\ope{\phi,2}$, which rescales all
Higgs couplings, dominates the most sensitive directions for all three
processes~\cite{madfisher}, typically followed by a combination of
$\ope{W}$ and $\ope{WW}$. Of the two $CP$-odd operators
$\ope{W\widetilde{W}}$ and $\ope{B\widetilde{B}}$, only the former can
be meaningfully constrained in these processes. In WBF Higgs
production, the sensitivity to this operator is best isolated in the
asymmetry $a_{\Delta \phi}$, which is not sensitive to any real
$CP$-even Wilson coefficients. But the corresponding Fisher
information still shows an admixture mixture of the imaginary Wilson
coefficient of the $CP$-even operator $\ope{WW}$, once again
demonstrating that additional theory assumptions are necessary to
measure $CP$ violation in this channel. In contrast, the genuine
$CP$-odd asymmetry $a_{\Delta \phi}$ in $ZH$ production is solely
sensitive to $CP$-violating operators, albeit at a reduced new physics
reach. As expected, the asymmetry $a_{\Delta E}$ is sensitive only to
the combination of $CP$ violation and absorptive physics modelled by
imaginary coefficients for the operator $\ope{W\widetilde{W}}$. In
both WBF and $ZH$ production, adding observables that measure the
momentum transfer significantly increases the information on all
operators, but at the cost of obfuscating the $CP$ interpretation of
the results. Finally, the Higgs decay is only really sensitive to the
combination of mostly $\ope{\phi,2}$ and $\ope{W}$ that affects the
total rate in this channel, and the physics reach in any other
direction in model space is severely hampered by the limited momentum
flow.

In the bottom panel we focus on the Fisher information on the
$CP$-violating Wilson coefficient $f_{W\widetilde{W}}$.  The grey bars
show the sensitivity assuming that all other considered operators are
zero, translated into the new physics reach on the right axis.  A
combination of the two leading WBF observables $p_{T, j_1}$ and
$\Delta \phi_{jj}$~\cite{phi_jj} captures almost the entire
phase-space information on $f_{W\widetilde{W}}$. When we profile over
arbitrary values of all of the $CP$-even parameters, including the
absorptive imaginary parts, this feature gets washed out, motivating a
multi-variate WBF analysis. The theoretically better-controlled $ZH$
production channel has a significantly smaller reach than the WBF
signature, but its reach for $f_{W\widetilde{W}}$ is literally
unaffected by other operators, thanks to the genuine $CP$-odd
observable. For the Higgs decay the only news which is worse than the
fact that there is very little information distributed over phase
space is that the genuine $CP$-odd asymmetry $a_{\Phi}$ is extremely
limited in reach.

Altogether, we find that a $CP$ measurement in WBF production
provides the best reach, but its interpretation is theoretically not
very clean.  A $CP$ measurement in $ZH$ production is less
model-dependent and more stable in terms of correlations, because we
can construct an appropriate genuine $CP$-odd observable. In both
cases, variables constructed to be sensitive to $CP$ can be combined
with information pertaining to the momentum transfer, which enhances
the effect of dimension-six operators compared to the Standard Model
amplitude. Finally, the Higgs decay is easily reconstructed and
analyzed, but has a very limited reach because of its limited momentum
transfer. Between the three processes we studied, there is no
unequivocally best signature to determine the $CP$ properties of the
Higgs-gauge sector, but there is clearly a worst.

\subsubsection*{Acknowledgments}

We would like to thank Kyle Cranmer, whose ideas laid the foundation
for our information-based approach to particle physics.  TMPT has
benefitted from conversations with Christoph Weniger.  JB is grateful
for the support of the Moore-Sloan Data Science Environment at
NYU. The work of FK and TMPT is supported in part by NSF under Grant
PHY-1620638. TP is supported by the DFG Forschergruppe \emph{New
  Physics at the LHC} (FOR~2239). The authors acknowledge support by
the state of Baden-W\"urttemberg through bwHPC.

\appendix
\section{Analytic form of the cross section}

The most general Lorentz structure of the Higgs-gauge couplings $H
V^\alpha (k_1) V^\beta (k_2)$ for on-shell gauge bosons can be written
as~\cite{hopkins,our_nelson,gino_mike}
\begin{align}
T^{\alpha\beta}
= \frac{2i}{v} \Big[ a_V m_V^2  g^{\alpha\beta} 
+ b_V(k_2^\alpha k_1^\beta-k_1\cdot k_2 g^{\alpha\beta})  
+ \beta_V  \epsilon^{\alpha\beta\gamma\delta} k_{1\gamma}k_{2\delta} \Big] \; .
\label{coup}
\end{align}
The SM at tree level is characterized by $a=1$ and
$b=\beta=0$. Radiative corrections in the SM or new light new
particles can introduce complex phases in $a,b$ and $\beta$. In the
Standard Model EFT, we can relate $a,b$ and $\beta$ to the Wilson
coefficients of the dimension-six operators,
\begin{align}
a_W=1 +m_W^2 \frac{f_W}{\Lambda^2}\qqquad
b_W=- m_W^2  \left( \frac{f_{W}}{\Lambda^2}  + 2\frac{f_{WW}}{\Lambda^2} \right) \qqquad
\beta_W= m_W^2 \frac{f_{W\widetilde{W}}} {\Lambda^2} \,,
\end{align}
and similar for the $HZZ$ couplings.  These couplings are constrained
by Higgs measurements. In particular, the decay mode $H \to \gamma
\gamma$ is sensitive to new physics contributions, since it first
appears at one-loop in the SM.  The dimension-six operators introduced
above induce an additional contributions described by the form factor
\begin{align}
b_\gamma = - m_W^2 s_W^2 \frac{f_{WW}+f_{BB}}{\Lambda^2}.
\end{align}
The absence of deviation from the SM in the diphoton channel therefore
implies $f_{BB}\approx-f_{WW}$.\bigskip

\begin{figure}[t]
  \centering	
  \includegraphics[width=0.4\textwidth]{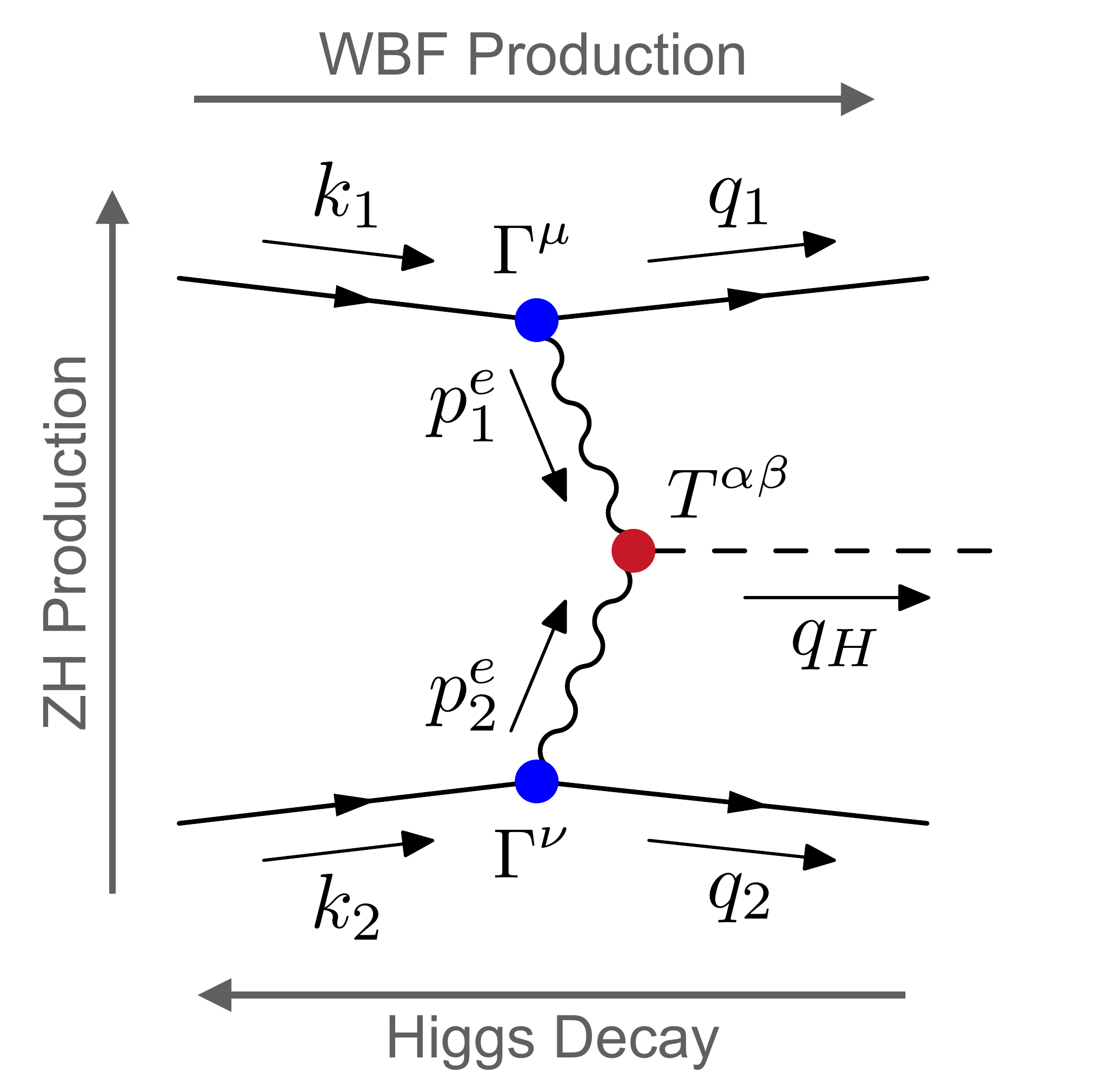}%
  \caption{Feynman diagram for WBF-Higgs production, $ZH$ production and Higgs decay to 4 leptons. The arrows indicate the momentum assignment used in the text.}
  \label{momenta}
\end{figure}

The three processes considered in this paper are at leading order
described by the single diagram shown in Fig.~\ref{momenta}. We
evaluate the corresponding matrix elements for the general form of the
Higgs coupling to vector bosons in Eq.\;\eqref{coup}, using the
$W$-boson mediated WBF Higgs production for illustration.  The other
processes shown in Fig.~\ref{fig:feyn} are related by appropriate
crossings.  In this case, the gauge-fermion coupling takes the form
$-i g/(2\sqrt{2}) \gamma^\mu (1-\gamma^5) V_{qq'}$
where $V_{qq'}$ is the CKM matrix.
It is instructive to introduce the combinations
\begin{align}
p_1^e = p_{11}-p_{12} \qquad 
p_1^o = p_{11}+p_{12} \qquad 
p_2^e = p_{21}-p_{22} \qquad 
p_2^o =p_{21}+p_{22} \; ,
\end{align}
where the $p_i^e$ correspond to the vector boson momenta. In cases of $ZH$ 
production and $H \to 4 \ell$ decay, the momenta $p_i^o$ will be odd under 
$C$-conjugation, while the $p_i^e$ are $C$-even. From these momenta, we construct 
eleven different Lorentz invariant observables describing the kinematics:
\begin{align}
\text{$C$-odd:}\qquad & C_1 = p_1^e \cdot p_2^o \qquad C_2 = p_2^e \cdot p_1^o \notag\\
\text{$C$-even:} \qquad&N_1=p_1^e\cdot p_2^e \qquad N_2=p_1^e\cdot p_1^e \qquad
                       N_3=p_1^o\cdot p_2^o \qquad N_4=p_2^e\cdot p_2^e \notag\\
\text{vanishing:}\qquad&V_1=p_1^e\cdot p_1^o=0 \qquad V_2=p_2^e\cdot p_2^o=0\notag\\
\text{$P$-odd:}\qquad& P_1 = \epsilon_{\alpha\beta\gamma\delta} p_1^{e\alpha} p_1^{o\beta} p_2^{e\gamma} p_2^{o\delta}
\end{align}
where we have used the fact that the fermions are approximately massless.
The squared matrix element for the process $ud \to du H$, averaged over initial spins and summed
over final spins, takes the form
\begin{align}
|\mathcal{M}|^2 &= \frac{g^4 |V_{ud}|^4 }{1024} \frac{N}{(p_1^2 - m_W^2)^2 (p_2^2 - m_W^2)^2} \notag \\
\text{with} \quad
N &=
|a|^2 f_{a} + |b|^2 f_{b} + |\beta|^2 f_{\beta} 
+ 2\text{Re}(ab^*) f^{R}_{ab}+ 2\text{Im}(ab^*) f^{I}_{ab} \notag\\
&\;\;\;+ 2\text{Re}(a\beta^*) f^{R}_{a\beta}+ 2\text{Im}(a\beta^*) f^{I}_{a\beta} 
+ 2\text{Re}(b\beta^*) f^{R}_{b\beta}+ 2\text{Im}(b\beta^*) f^{I}_{b\beta} \; .
\end{align}
The individual contributions are
\begin{align}
f_{a} &= m_V^4 ((N_1+N_3)^2-(C_1+C_2)^2)\notag\\
f_{b} &=  (C_1 C_2 - N_1 N_3 )^2 + N_2 N_4 (C_1^2 + C_2^2 + 2 N_1 N_3) + N_2^2 N_4^2 \notag\\
f_{\beta} &=  N_2 N_4 (2 C_1 C_2 +N_1^2 + N_3^2)-(C_1 C_2 - N_1 N_3 )^2 -  N_2^2 N_4^2\notag\\
f_{ab}^R &=  m_V^2  (N_1+N_3) (C_1 C_2 - N_1 N_3 - N_2 N_4)\notag\\
f_{ab}^I  &=  m_V^2 P_1 (C_1+C_2)\notag\\
f_{a\beta}^R &= -  m_V^2 P_1 (N_1+N_3) \notag\\
f_{a\beta}^I  &=  m_V^2 (C_1+C_2) (N_1 N_3 + N_2 N_4 - C_1 C_2)\notag\\
f_{b\beta}^R &=  P_1 (N_1 N_3 + N_2 N_4 - C_1 C_2) \notag\\
f_{b\beta}^I  &=  (C_1+C_2) N_2 N_4 (N_1+N_3)  \; .
\end{align}
For a non-vanishing expectation value of the $P$-odd observable $P_1$,
the squared matrix element must contain a term linear in $P_1$. Such
terms are generated either in the presence of $CP$-violating new
physics, when either $\text{Re}(a\beta^*)\neq0$ or
$\text{Re}(b\beta^*)\neq0$, or in the presence of the absorptive phase
$\text{Im}(ab)\neq0$.

In the processes involving a $Z$ boson, the process may also enjoy a
well defined transformation under charge conjugation. In this case, a
non-vanishing expectation value for the $C$-odd observables $C_{1,2}$
requires the squared matrix element to contain a term linear in
$C_{1,2}$. Such terms are generated only via rescattering effects when
$\text{Im}(ab^*),\text{Im}(a\beta^*),\text{Im}(b\beta^*) \neq 0$.  The
measurement of $C_i$ sometimes requires the identification of fermion
charges, and therefore is not possible for all processes.  Note that
$f_{ab}^I$ is both $C$-odd and $P$-odd and therefore $CP$-even. Thus,
it does not contribute to a non-vanishing expectation value of both
$P$ and $C_i$ if the initial state is $CP$-symmetric, verifying the
observation that absorptive phases can induce an asymmetry in $P_1$ in
WBF Higgs production, but not in $ZH$ production or $H \to 4\ell$
decay.

\end{fmffile}


\end{document}